\documentclass[usenatbib]{mnras}

\usepackage{graphicx}	
\usepackage{amsmath}	
\usepackage{amssymb}	

\usepackage[flushleft]{threeparttable}

\usepackage{Custom_Functions}

\usepackage[usenames, dvipsnames]{color}
\definecolor{mycolor}{RGB}{0,0,0}
\definecolor{newaddcolor}{RGB}{255,0,255}

\usepackage[normalem]{ulem}

\title[$\Gamma_{\rm HI}$ at $z<0.5$]{Intergalactic Lyman continuum photon budget in the past 5 billion years}

\author[Gaikwad et.al]{Prakash Gaikwad$^{1}$\thanks{E-mail: \href{prakashg@ncra.tifr.res.in}{prakashg@ncra.tifr.res.in}}, 
Vikram Khaire$^{1,2}$,
Tirthankar Roy Choudhury$^{1}$,
\newauthor{and Raghunathan Srianand$^{2}$}
\\
$^{1}$National Centre for Radio Astrophysics, Tata Institute of Fundamental Research, Pune 411007, India \\
$^{2}$Inter-University Centre for Astronomy and Astrophysics (IUCAA), Post Bag 4, Pune 411007, India}

\date{}

\pubyear{2015}

\begin{document}
\label{firstpage}
\pagerange{\pageref{firstpage}--\pageref{lastpage}}
\maketitle


\begin{abstract}
We constrain the H~{\sc i} photoionization rate $(\Gamma_{\rm HI})$ at $z \lesssim 0.45$ by comparing the flux probability distribution function and power spectrum of the Ly$\alpha$ forest data along 82 QSO sightlines obtained using Cosmic Origins Spectrograph with models generated from smoothed particle hydrodynamic simulations. We have developed a module named ``Code for Ionization and Temperature Evolution ({\sc cite})'' for calculating the intergalactic medium (IGM) temperature evolution from high to low redshifts by post-processing the {\sc gadget-2} simulation outputs. Our method, that produces results consistent with other simulations, is computationally less expensive thus allowing us to explore a large parameter space. It also allows rigorous estimation of the error covariance matrix for various statistical quantities of interest. We find that the best-fit $\Gamma_{\rm HI}(z)$ increases with $z$ and follows $(4 \pm 0.1) \times 10^{-14}\:(1+z)^{4.99 \pm 0.12}$ s$^{-1}$. At any given $z$ the typical uncertainties $\Delta \Gamma_{\rm HI} / \Gamma_{\rm HI}$ are $\sim 25$ per cent which contains not only the statistical errors but also those arising from  possible degeneracy with the thermal history of the IGM and cosmological parameters and uncertainties in fitting the QSO continuum. These values of $\Gamma_{\rm HI}$ favour the scenario where only QSOs contribute to the ionizing background at $z<2$.
Our derived $3\sigma$ upper limit on average escape fraction is $0.008$, consistent with measurements of low-$z$ galaxies.
\end{abstract}


\section{Introduction}
\label{sec:intrduction}

It is believed that the H~{\sc i} gas in the intergalactic medium (IGM) got reionized at $z\ge5.5$ \citep{madau1999,becker2001,fan2001,fan2006,robertson2010,bolton2011,planck2015,kstp2015}. At subsequent epochs the IGM  is maintained at a highly ionized state by ultra-violet background (UVB) radiation ($\lambda \le 912 \: {\rm \AA}$).
This UVB is contributed by radiation from blackhole accretion in Quasi Stellar Objects
(hereafter QSOs) and  stellar light escaping from galaxies \citep[see for example,][]{miralda1990,HM1996,fardal1998,shull1999}. 
The stellar contribution to UVB depends crucially on the fraction of ionizing photons escaping the galaxies known as the escape fraction $(f_{\rm esc})$. 
 The $f_{\rm esc}$, in principle, depends on various physical factors such as the galaxy mass, morphology,  composition of the interstellar medium (ISM),  spatial distribution of gas and supernova rates \citep{ricotti2000,gnedin2008,fernandez2011,benson2013,kim2013,cen2015,roy2015}.  
As a result, there is no consensus among different models of $f_{\rm esc}$. 
Measuring $f_{\rm esc}$ directly from observations too is quite challenging.
The reported values of $f_{\rm esc}$ at $2\le z\le 4$ vary between 0.01 to 0.2  \citep{iwata2009,boutsia2011,nestor2013,cooke2014,siana2015,mostardi2015,micheva2015,grazian2016,smith2016,vasei2016}.
At $z < 2$, apart from the detection of high $f_{\rm esc}$ in few individual galaxies \citep{deharveng2001,bergvall2006,leitet2013,borthakur2014,izotove2016,leitherer2016}, the $3\sigma$ upper limits on average $f_{\rm esc}$ obtained by stacking samples of galaxies is $\leq 0.02$ \citep{cowie2009,bridge2010,siana2010,rutkowski2015}.

An alternate way of constraining $f_{\rm esc}$ (and hence the stellar contribution to the UVB) is by measuring H~{\sc i} photoionization rate $(\Gamma_{\rm HI})$ \citep[see ][]{inoue2006,kstp2015}.
The $f_{\rm esc}$ also determines the shape of the UVB \citep{khaire2013} which is important for modeling the distribution of ions in the IGM detected in the QSO spectra \citep{shull2014,rahmati2016,finn2016,oppenheimer2016}. 
The UVB estimated at any given redshift $z_0$ depends on emissivities of radiating sources  and the IGM opacity (contributed mainly by the Lyman Limit Systems with $N_{\rm HI} \ge 10^{17}$ cm$^{-2}$) over the large redshift range $z \geq z_0$. 
Therefore, measurements of  $\Gamma_{\rm HI}(z_0)$, can be useful to constrain the IGM opacity evolution at $z \geq z_0$, especially at low-$z$ where it is ill-constrained.

$\Gamma_{\rm HI}$ is usually constrained in the literature using three methods: (i) The first method uses the H~{\sc i} absorption in the proximity of QSOs \citep{bajtlik1988,scott2000,kulkarni1993,srianand1996,dallaglio2009,calverley2011}. 
The main uncertainties in this method arise because of the anisotropies in the QSO emission \citep{schirber2004,kirkman2008} and possible density enhancements around the QSO host galaxies \citep{rollinde2005,Guimaraes2007,FG2008}.
(ii) The second method, mainly useful at low-$z$, is based on the measured H$\alpha$ surface brightness from the outskirts of nearby galaxies ($z \sim 0$) and high velocity cloud at the edges of our galaxy \citep{kutyrev1989,songaila1989,vogel1995,madsen2001,weymann2001,adams2011}. However this measurement too is uncertain because of the assumptions made about the geometries of the H$\alpha$ emitting gas \citep[see ][]{shull2014}.
(iii) The third method of constraining $\Gamma_{\rm HI}$ is by simulating the observed properties of the Ly-$\alpha$ forest (far away from the proximity of QSOs) such as the H~{\sc i} column density distribution function (CDDF) \citep{kollmeier2014,shull2015}, the flux probability distribution function (PDF) and the flux power spectrum (PS), which forms the basis of the analysis presented in this paper.

The basic idea behind using the Ly-$\alpha$ forest to constrain $\Gamma_{\rm HI}$ is that, under the fluctuating Gunn-Peterson approximation \citep[hereafter FGPA; see][]{FGPA,weinberg1998,croft1998}, the Ly-$\alpha$ optical depth scales as $\Gamma^{-1}_{\rm HI}$ and hence can be used to constrain it. Observed statistical properties of the Ly-$\alpha$ forest are compared with those from an appropriate model with $\Gamma_{\rm HI}$ as one of the free parameters. The other free parameters in these models are those describing the thermal history of the IGM and the cosmological parameters which can be degenerate with $\Gamma_{\rm HI}$.
To estimate the uncertainty in $\Gamma_{\rm HI}$ at $2 \leq z \leq 4$ due to its degeneracy with other parameters, \citet{bolton2005,bolton2007,faucher2008c} used scaling relations derived from their hydrodynamical  simulations \citep[Table 4 in][]{bolton2007}.  
At these redshifts, it is well known from numerical simulations that the Ly-$\alpha$ forest arises from the low-density diffuse medium which accounts for $90$ per cent of the baryons \citep{cen1994,zhang1994,hernquist1996,miralda1996,rauch1997,paschos2005,faucher2008,bolton2009}. Hence a relatively simple model \citep{bi1997,trc2001} is sufficient to reliably constrain $\Gamma_{\rm HI}$. On the other hand at low-$z$ ($z \leq 1.6$),  only a small fraction ($\sim 30 - 40$ per cent) of the baryons are in diffuse medium responsible for Ly-$\alpha$ forest \citep{theuns1998a,dave2001b,dave2010,smith2011,tepper2012,shull2015}. It turns out that a significant fraction ($\sim 30 - 50$ per cent) of the baryons are in a phase known as the warm hot intergalactic medium (WHIM) \citep{cen1999,dave2001a,cen2006a,cen2006b,lehner2007,dave2010,smith2011,shull2012b} and they are difficult to detect in either emission or absorption in the UV/optical bands.
Hence to measure $\Gamma_{\rm HI}$ at low-$z$, one needs simulation incorporating all these effects. 
Furthermore there is also a strong possibility that  SNe and AGN feedback processes can inject thermal energy into the IGM which may change the density-temperature distribution (i.e the phase diagram) of the baryons \citep{dave2010,smith2011,shull2015,tepper2012}, thus probably affecting the Ly-$\alpha$ forest observable. 

To observe the low-$z$  ($z \leq 1.6$) Ly-$\alpha$ forest, one needs the UV spectrograph onboard space based telescope.
Thanks to a large survey using the Cosmic Origins Spectrograph (COS) onboard the Hubble Space Telescope (HST) \citep{danforth2016}, there are now constraints on $\Gamma_{\rm HI}$, e.g., by using column density distribution of low-$z$ Ly-$\alpha$ forest \citep{kollmeier2014,shull2015}, and by modeling the observed metal abundances of ions \citep{shull2014} using {\sc cloudy} \citep{ferland1998}. It turns out that there is a tension between the inferred $\Gamma_{\rm HI}$ by \citet{kollmeier2014} and \citet{shull2015}, where both use Ly-$\alpha$ forest data by \citet{danforth2016} but different simulations. The inferred $\Gamma_{\rm HI}$ values disagree by a factor of $\sim 2.5$. Given such a wide disagreement, it is worth taking an independent closer look at the $\Gamma_{\rm HI}$ measurements at low-$z$ using the Ly-$\alpha$ forest, in particular a careful analysis of the systematics in the data as well as modeling uncertainties.

The main aim of this paper is to measure $\Gamma_{\rm HI}$ from the Ly-$\alpha$ forest data by \citet{danforth2016}, using two different statistics, namely the flux PDF and the flux PS that are regularly used in high-$z$ studies. The method is based on performing smooth particle hydrodynamical (SPH) simulations to generate the density and velocity distributions of baryons and then post-processing the outputs to solve for the IGM temperature in presence of a UVB. For this purpose, we have developed a module, called the ``Code for Ionization and Temperature Evolution'' ({\sc cite}), to evolve the IGM temperature from high-$z$ to the low-$z$ of our interest. The advantage of this method is that it is computationally less expensive and sufficiently flexible to account for variations in the thermal history. Our analysis allows us to study the degeneracy between $\Gamma_{\rm HI}$ and parameters related to the thermal history. The other significant step in our analysis is that we calculate the errors on $\Gamma_{\rm HI}$, unlike \citet{shull2015,kollmeier2014}, by estimating the error covariance matrix from the simulations using a method similar to \citet{rollinde}, thus avoiding any non-convergence that may arise from the limited sample of the observed data.

The plan of this paper is as follows: The observational data used in our analysis are discussed in \S\ref{sec:observation}. Details of the simulations, along with our method of calculating the temperature evolution, is discussed in \S\ref{sec:simulation}. The two statistics used in this paper (i.e., the flux PDF and PS) and the associated errors are discussed in \S\ref{sec:method}. The main results of our work are discussed in \S\ref{sec:results}, where we match the simulations with the observed data to constrain $\Gamma_{\rm HI}$. We also discuss the various statistical and systematic uncertainties in the measured $\Gamma_{\rm HI}$, and show the consistency of our derived $\Gamma_{\rm HI}$ by comparing the H~{\sc i} column density distribution from observation with simulation. Finally we use the ionizing background computed by \citet{ks15a} using the updated emissivities and IGM opacities  to constrain $f_{\rm esc}$ from the evolution of $\Gamma_{\rm HI}$. 
We summarize our findings in \S\ref{sec:summary}.  We use the flat $\Lambda$CDM cosmology with parameters $(\Omega_{\Lambda},\Omega_{m},\Omega_{b},\sigma_8,n_s, h,Y) = (0.69,0.31,0.0486,0.83,0.96,0.674,0.24)$ \citep{planck2015}. All the distances are in comoving Mpc unless and otherwise mentioned. We use $\Gamma_{\rm 12}$ to express $\Gamma_{\rm HI}$ in units of $10^{-12}$ s$^{-1}$.


\section{HST-COS QSO absorption spectra}
\label{sec:observation}

We used the publicly available data from a survey\footnote{https://archive.stsci.edu/prepds/igm/} of low redshift Ly-$\alpha$ performed by \citet{danforth2016} using HST-COS. The sample consists of 82 UV-bright QSO sightlines, with the QSOs being distributed across the redshift range $0.0628$ to $0.852$. The observations were carried out between July 2009 and August 2013. Using accurate data reduction process and careful subtraction of the background, \citet{danforth2016} have produced high signal to noise ratio (SNR) Ly-$\alpha$ forest spectra in the observed wavelength range $1100 {\rm \AA}$ to $1800 {\rm \AA}$. In addition, \citet{danforth2016} have fitted the continuum and identified several thousand absorption features using a semi-automated procedure. These absorption features arise not only from H~{\sc i} Lyman series lines from the IGM  but also from other intervening absorbers and from the Galactic interstellar medium. The redshift range for Ly-$\alpha$ lines covered by each sightline is shown in Fig.\ref{fig:redshift-coverage}. The sharp cutoff shown by blue curly bracket in the figure is because of the limited wavelength range covered by the spectrograph.  We assume that a region of comoving size of up to $25 h^{-1}$ cMpc around a QSO can be affected by the proximity effect of the QSO itself \citep{lidz2007}, hence we exclude the corresponding section blueward of the Ly-$\alpha$ emission line. 

\InputFig{Redshift_Coverage.pdf}{80}
{The redshift range covered by the Ly-$\alpha$ forest for the 82 HST-COS spectra used in this work (see \S\ref{sec:observation}). The vertical dashed lines show the redshift bins with centers at $z =0.1125,0.2,0.3,0.4$ and width $\Delta z = 0.075,0.1,0.1,0.1$ respectively. The redshift bins used in this work are shown by roman numerals. The sharp cutoff shown by blue curly bracket is arising from the red wavelength cutoff of the COS-160M grism used (at $z=0.48$). In these cases the COS spectra do not cover the Ly-$\alpha$ emission from the QSOs.}
{\label{fig:redshift-coverage}}

We divide the data into 4 different redshift bins.  Three bins are centered on $z = 0.2 ,0.3, 0.4$ with a width of $\Delta z = 0.1$. 
We chose the lowest redshift bin $z = 0.1125 $ with $ \Delta z = 0.0375$ to avoid the contamination from the foreground geo-coronal line emission at $z<0.075$.
We are then left with Ly-$\alpha$ absorption from $50,31,16,12$ sightlines each in the redshift bins with $z=0.1125,0.2,0.3,0.4$, respectively. The redshift bins chosen for analysis are indicated by roman numerals I, II, III and IV in Fig. \ref{fig:redshift-coverage}. 
The values of the SNR for the Ly-$\alpha$ forest spectra vary between $5$ and $17$. Each observed spectrum has a resolution of $\sim 17$ km s$^{-1}$.
Table \ref{tab:obs-property} summarizes the properties of the observed Ly-$\alpha$ forest data in the four identified redshift bins.
The table contains the redshift range of observation ($z_{\rm obs}$), the redshift of simulation box used ($\overline{z}_{\rm sim}$) for comparison, number of sightlines used and SNR range for each bin. 
A sample observed spectrum (towards the QSO 3C57) is shown in the top panel of Fig. \ref{fig:obs-sim-spectra}. In addition to the Ly-$\alpha$ absorption, the spectrum contains higher Lyman-series and various metal absorption lines as shown in the figure. 
We fit these lines with gaussian and replace them with appropriate continuum added with gaussian random noise (with the same SNR) as shown in  middle panel of Fig. \ref{fig:obs-sim-spectra}. 
We use these clean spectra (i.e., metal line and higher order Ly-series line removed) to match observations with simulations using flux PDF and flux PS.

\begin{table}
\centering
\caption{Details of the HST-COS data used in different redshift bins. ($\overline{z}_{\rm sim}$ is redshift of simulation box used for comparison)}
\begin{tabular}{ccccc}
\hline
Redshift & $z_{\rm obs}$ & $\overline{z}_{\rm sim}$ &Number  & SNR \\
bin    & & &of QSOs & Range \\ \hline
I & 0.075 - 0.15 & 0.1 & 50 &  14.5 - 16.9 \\
II & 0.15 - 0.25  & 0.2 & 31 & 13.0 - 14.4 \\ 
III & 0.25 - 0.35 & 0.3 & 16 & 6.3 - 13.3 \\ 
IV & 0.35 - 0.45 & 0.4 & 12 & 5.8 - 6.9 \\ \hline
\end{tabular}
\label{tab:obs-property}
\end{table}

\InputFigCombine{observed_simulated_spectra.pdf}{160}
{\emph{Top panel} shows the observed Ly-$\alpha$ forest towards the QSO 3C57. H~{\sc i} Ly series and metal lines as identified by \citet{danforth2016} are also marked.
\emph{Middle panel} shows the same spectrum after these lines are removed and replaced by a continuum added with random noise with the same SNR as in the original spectrum (see \S\ref{sec:observation}).
\emph{Bottom panel} shows the simulated spectrum towards a random line of sight in our simulation box. 
The simulated spectrum is convolved with the appropriate line spread function of HST-COS and added with noise having SNR similar to that of 3C57 (see \S\ref{subsec:ly-alpha-forest}).
}
{\label{fig:obs-sim-spectra}}

\section{Details of our simulations}
\label{sec:simulation}

We generate the cosmological density and velocity fields using the smoothed particle hydrodynamic code {\sc gadget-2}\footnote{http://www.mpa.mpa-garching.mpg.de/gadget/} \citep{springel2005}. The initial conditions for the simulations are generated at a redshift $z=99$ using the publicly available code {\sc 2lpt}\footnote{http://cosmo.nyu.edu/roman/2LPT/} \citep{2lpt2012}. We use $1/30^{th}$ of the mean inter-particle distance as our gravitational softening length. The simulation outputs are stored at a redshift interval of 0.1 between $z = 2.1$ and $z=0$. We use $2$ simulation boxes containing $512^3$ dark matter and an equal number of gas particles in a cubical  box. Both simulation boxes are $50 h^{-1}$ cMpc in size with different initial conditions. We use these boxes to study cosmic variance. 
As shown by \citet{smith2011}, the phase distribution of baryons in simulation box at low-$z$ is converged if the box size $50 h^{-1}$ cMpc or above. 
The simulations used in this work do not include AGN feedback, outflows in the form of galactic wind or micro-turbulence. Note that the gas heating due to hydrodynamical processes arising from structure formation is incorporated in {\sc gadget-2}. However radiative heating and cooling processes are not incorporated in {\sc gadget-2} and we include them in post processing step (see \S  \ref{sec:equation-of-state}). 
We use {\sc gadget-2} with post-processing instead of {\sc gadget-3} in order to probe the wide range of parameter space. Later we compare our results with those obtained using different simulations in the literature.
 
\subsection{Density, velocity and temperature} 
\label{subsec:dens-vel-temp}
Each {\sc gadget-2} output snapshot contains the position, velocity, internal energy per unit mass and smoothing length $l$ of each smooth particle hydrodynamic (SPH) particle. We evaluate the above quantities on a uniform grid in the box using the smoothing kernel \citep{springel2005},
\begin{equation} \label{eq:smth-kernal}
    W(r,l) = \frac{8}{\pi \: l^3}
\begin{cases}
    \;\; 1 - 6 \bigg( \frac{r}{l} \bigg)^2 + 6\bigg( \frac{r}{l} \bigg)^3,&  0 \leq \frac{r}{l} \leq \frac{1}{2}\\
    \;\; 2\bigg( 1 - \frac{r}{l} \bigg)^3, & \frac{1}{2} \leq \frac{r}{l} \leq 1 \\ 
    \;\; 0, & \frac{r}{l} > 1
\end{cases}
\end{equation}
where $r$ is the distance between the grid point and the particle position. 
The density at the $i^{th}$ grid is simply the sum of density contribution from all particle weighted by the smoothing kernel
\begin{equation} \label{eq:gadget-density}
\rho_i = \sum \limits_{j=1}^{N} \; m_j \;W(|\textbf{r}_{ij}|,l_j),
\end{equation}
where $N$ is the total number of particles, $|\textbf{r}_{ij}|$ is the distance between the $i^{th}$ grid point and $j^{th}$ particle. $m_j$ and $l_j$ are the mass and smoothing length of the $j^{th}$ particle respectively. The overdensity at the grid point $i$ is given by
\begin{equation} \label{eq:gadget-overdensity}
 \Delta_i = \frac{\rho_i}{\overline{\rho}}, \;\;\;\; \text{with} \;\;\;\; \overline{\rho} = \frac{1}{N} \sum \limits_{i=1}^{N} \rho_i
\end{equation}
where $\overline{\rho}$ is the average density. In this work the symbol $\Delta$ is used for baryon overdensity.
The density weighted estimate of any quantity $f$ at the $i^{th}$ grid point is given by
\begin{equation}\label{eq:gadget-weigthing}
f_i = \frac{\sum \limits_{j=1}^{N} \; f_j \; m_j \;W(|\textbf{r}_{ij}|,l_j)}{\sum \limits_{j=1}^{N} \; m_j \;W(|\textbf{r}_{ij}|,l_j)}
\end{equation}
where $f_j$ is the value of the quantity for the $j^{th}$ particle. The quantity $f$ could be one of the velocity components ($v_x$, $v_y$, $v_z$) or the internal energy per unit mass $u$.
The temperature of the gas as computed by {\sc gadget-2}, which we denote as $T_g$, can be obtained from the internal energy using the relation
\begin{equation}\label{eq:gadget-temperature}
T_g = \frac{2 m_p}{3 k_B} \; u,
\end{equation}
here we have assumed a monoatomic gas composition with the ratio of specific heats given by $5/3$. In the above expression, $m_p$ is the mass of a proton and $k_B$ is the Boltzmann's constant.
 
Fig. \ref{fig:simulation-slice} shows a two-dimensional slice from our simulation box at $z = 0.3$.
The left-hand panel and middle panel shows, respectively, the overdensity $(\Delta)$ and the temperature $(T_g)$ of baryons as obtained from the {\sc gadget-2} output snapshots.  By comparing the left-hand and middle panels, one can see that the temperature distribution broadly traces the density distribution. This is related to the fact that the high density regions are heated because of hydrodynamical processes. Hereafter we will refer to this heating simply as shock heating. The temperature can be as high as $\sim 10^7$ K in the vicinity of collapsed objects. The voids, on the other hand, remain extremely cool at temperatures $\sim 10$ K.

However, the above scenario does not capture all the relevant physics, in particular the photoheating of the low-density gas by UVB to higher temperatures and various cooling processes. It is indeed found from other simulations, that take into account the additional heating and cooling processes \citep{huignedin,mcdonaldlinefit}, that the temperature and density follow a reasonably tight relation (which we will refer to as the $T-\Delta$ relation) for mildly non-linear densities, i.e.  $\Delta \leq 10$. 
As {\sc gadget-2} does not include processes like the photoheating and radiative cooling, we find that the resulting temperature ($T_g$) and density relation does not show any power-law correlation. 

This shortcoming has been addressed in subsequent versions of the {\sc gadget}, e.g., {\sc gadget-3} \citep[as discussed in ][]{bolton2006} where one can perform the simulation in presence of a UVB. In this work, however, we follow a slightly different approach to account for the effects of photoionizing UVB and radiative cooling. Our method involves post-processing the {\sc gadget-2} output to calculate the temperatures. We will show that this method produces results which are consistent with other works. The advantage of our method, however, is that we are able to explore the parameter space more efficiently without having to perform the full SPH simulation multiple times.
In the following section we outline our method to evolve the IGM temperature in the post-processing step using `Code for Ionization and Temperature Evolution' ({\sc cite}). 

\InputFigCombine{Slice_1.pdf}{175} {%
Two-dimensional slices of width 0.1 $h^{-1}$ cMpc  obtained from the {\sc gadget-2} output snapshot at $z=0.3$. \emph{Left-hand panel:} the distribution of baryon overdensity $\Delta$. Color scheme is such that red and blue color represent highest density and lowest density regions respectively. \emph{Middle panel:} the gas temperature $T_g$ from {\sc gadget-2} (see \S\ref{subsec:dens-vel-temp}). \emph{Right-hand panel:} the gas temperature $T$ predicted after evolving the temperature from $z_1=2.1$ (initially at $z_1$, $T_0 = 15000$ K and $\gamma=1.3$) using our post-processing module {\sc cite}  (see \S\ref{subsec:cite}). The highly overdense regions are at higher temperatures because of the shock heating resulting  from the structure formation. The color scheme in middle and right-hand panel is such that red and blue color corresponds to highest temperature and lowest temperature regions respectively.}
{\label{fig:simulation-slice}}

\subsection{Code for Ionization and Temperature Evolution ({\sc cite})}
\label{subsec:cite}

\label{sec:equation-of-state}

The temperature evolution equation  for an overdense region in the IGM is given by \citep{huignedin},
\begin{eqnarray}\label{eq:temperature-evolution}
\frac{dT}{dt} &=& \left(-2HT + \frac{2T}{3\Delta} \: \frac{d \Delta}{dt} +  \frac{dT_{shock}}{dt} \right) 
\nonumber\\
&+&  \frac{T}{\sum_i X_i} \: \frac{d \sum_i X_i}{dt} + \frac{2}{3 \: k_B \: n_b} \frac{dQ}{dt}.
\end{eqnarray}
In the above equation the first three terms on right hand side (i.e., those in the large parenthesis) represent, respectively, the rate of cooling due to Hubble expansion, adiabatic heating and/or cooling arising from the evolution of the densities of gas particles and the change in temperature because of shocks which can be an important source of heating at low redshifts \citep{dave2001a,dave2001b}. These three mechanisms are taken into account in the default run of the {\sc gadget-2}. The fourth term on the right hand side represents the change in internal energy per particle arising from the change in the number of particles.
The last term accounts for other heating and cooling processes, e.g., photo-heating and radiative cooling. The radiative cooling processes can, in principle, include cooling from recombinations, collisional ionization, collisional excitation, inverse Compton scattering and free-free emission.

As we discussed earlier, the default run of {\sc gadget-2} results in temperatures that are too low at low densities, and does not show the tight $T - \Delta$ correlation at low to moderate overdensities. Therefore it is important to incorporate the effects of photoheating arising from the UVB to rectify the two problems. The method we follow to account for the photoheating and radiative cooling is as follows:

\begin{enumerate}
\item \label{step-init} We start with the output snapshots at a moderately high redshift, in our case it is taken to be $z_1 = 2.1$. This is an optimum redshift for our purpose as the He~{\sc ii} reionization is likely to be completed by then \citep{kriss2001,theuns2002,zheng2004,shull2004,khaire2013,worseck2014} and thus the ionizing radiation can be taken to be uniform. If \emph{all} the gas particles at $z_1$ follow a power-law equation of state, then the temperature would be given by\footnote{We varied the $\Delta$ cutoff in Eq. \ref{eq:power-law-T-D} from $10$ to $5$ and $15$ and found that the resulting $T-\Delta$ relations at low-$z$ are not sensitive to our choice of this cutoff.}
\begin{equation}\label{eq:power-law-T-D}
  \begin{aligned}
    T_1 \equiv T(z_1) &= T_0 \; \Delta^{\gamma-1} \; \mbox{ for } \; \Delta < 10 \\
        &= T_0 \; 10^{\gamma-1} \; \mbox{ for } \; \Delta \geq 10,
  \end{aligned}  
\end{equation}
where $T_0$ and $\gamma$ are free parameters. We have assumed that the high density gas with  $\Delta \geq 10$ is able to cool via atomic processes and hence have temperatures smaller than what is implied by the power-law \citep{theuns1998b}. Note that the temperature obtained using Eq. \ref{eq:power-law-T-D}, in general, will be different from that obtained from the {\sc gadget-2} output which we denote as $T_{g,1} \equiv T_g(z_1)$.

To obtain the \emph{actual} temperature of a gas particle, we use the following argument: If $T_{g,1} > T_1$ for that particle, then it may have been shock heated in a recent time step, and hence must have moved away from the $T-\Delta$ relation. In that case the particle temperature is taken to be $T_{g,1}$. Otherwise we assume the particle temperature to be following the equation of state and assign it as $T_1$.

We calculate the initial fractions of different ionized species (i.e., fraction of all ionization states of hydrogen and helium and hence the fraction of free electrons) by assuming ionization equilibrium to hold at the initial redshift $z_1 = 2.1$, which is a reasonable approximation for optically thin gas in the post-He~{\sc ii} reionization era \citep{bolton2008,becker2011}.  At this redshift we used $\Gamma_{\rm HI}$ consistent with QSO dominated ($f_{\rm esc} = 0$)  \citet{khairepuc} (hereafter KS15) UVB.

\item \label{step-start} Given the initial temperature and the ionization fractions, it is straightforward to calculate the last two terms on the right hand side of Eq. \ref{eq:temperature-evolution}. The expressions for the heating and cooling rates used are taken from \citet{theuns1998b} \citep[for similar expressions see][]{sutherland1993, katz1996, weinberg1997, wiersma2009}.
To estimate the photo-heating rate at any given $z$, we used QSO dominated (i.e. $f_{\rm esc} = 0$) KS15 UVB model.

\item To obtain the particle temperature at the next redshift $z_2 = z_1 - \Delta z$, we first compare the {\sc gadget-2} temperatures $T_{g,1}$ and $T_{g,2}$ at the two redshifts and thereby check whether the gas particle is shock heated within that time interval.
If for a particle $T_{g,2} < T_{g,1}$ then the particle is not shock heated. 
In this case we solve the Eq. \ref{eq:temperature-evolution} by neglecting third term (i.e., the one corresponding to the shock heating) on the right hand side. However if the particle is shock heated\footnote{We self consistently check if the particle is shock heated by solving Eq. \ref{eq:temperature-evolution} for {\sc gadget-2} temperatures i.e., by using only the terms in the parenthesis on right hand side of Eq. \ref{eq:temperature-evolution}.} $T_{g,2} > T_{g,1}$ then we solve the same equation taking into account all the terms.
\item  \label{step-stop} For redshift $z_2$ we solve \emph{non-equilibrium} ionization evolution equation to calculate various ionization fractions. In addition to the photoionization, we also include the collisional ionization in the non-equilibrium ionization evolution equation.
 \item We then repeat the step \ref{step-start} to \ref{step-stop} for subsequent redshifts and evolve the temperature of all the gas particles to our desired redshift.
\end{enumerate}
Since the differential Eq. \ref{eq:temperature-evolution} is ``stiff'', it tends to be numerically unstable if the time-step between two snapshots is too large.
To circumvent such difficulties,  we divide the time-step between  two neighboring redshifts  into 100 smaller steps. We linearly interpolate the {\sc gadget-2} temperature and densities for these intermediate time-steps. We have checked the effect of varying number of time-steps on $T-\Delta$ relation and found that the results converge as long as the number of intermediate steps is 50 or more. We incorporated the above method in a module called Code for Ionization and Temperature Evolution  `{\sc cite}'.

The \emph{middle} and \emph{right-hand} panels in Fig. \ref{fig:simulation-slice} show the comparison between {\sc gadget-2} temperature $(T_g)$ and temperature from {\sc cite} $(T_c)$. Note that the temperature scales in middle and right-hand panels are different. The highest temperature due to hydrodynamical processes is nearly same for $T_g$ and $T_c$. But the lower temperature scale is considerably different due to additional processes incorporated through {\sc cite}.
On an average, $T_c$ also traces the density field shown in \emph{left} hand panel of Fig. \ref{fig:simulation-slice}. 

Fig. \ref{fig:eos-T-15000-G-1-3} shows the resulting distribution at $z = 0.3$ in the $T - \Delta$ plane, which is often called as the ``phase diagram''. Note that the temperature and density plotted in the figure are volume-averaged, i.e., they are calculated by using the SPH kernel (see Eq. \ref{eq:gadget-weigthing}). For the plot we shoot 20000 random lines of sight through simulation box and calculate temperature and density on the grid points. The initial values at $z_1 = 2.1$ are chosen to be  $T_0=15000 \; {\rm K}$ and $\gamma=1.3$ (we refer to these values as initial $T_0$ and $\gamma$), the corresponding equation of state is shown by the red dashed line. The color coding represents density of points in logarithmic scale (i.e., the red color represents highest density of points). We can see from this figure that our simulation using {\sc cite} is able to produce the equation of state at low and moderate overdensities. In fact most of the grid points follow a power-law $T-\Delta$ relation (black dashed line) described by $T_0=4902 \; {\rm K}$ and $\gamma=1.53$  (at $z=0.3$) which is consistent with results from other low-$z$ hydrodynamical simulations \citep{dave2001b,dave2010,smith2011,shull2012b,tepper2012,shull2015}. It is interesting to note that a significant fraction of points are at very high temperatures $T > 10^5 \; {\rm K}$ forming WHIM. 
We defer a detailed comparison of our simulations with those available in the literature to \S \ref{subsec:comparison-other-simulation}

\InputFig{T-15000-G-1-3.pdf}{80}
{Distribution of grid points in the $T - \Delta$ plane at $z=0.3$ when the temperatures are estimated using {\sc cite} (see \S\ref{subsec:cite} for details of {\sc cite}). The color scale indicates the density of points are shown (in logarithmic scale). At the initial redshift $z_1=2.1$ the values of the free parameters are chosen as $T_0=15000 \: {\rm K} $ and $\gamma=1.3$ (model $T15-\gamma1.3$ in Table \ref{tab:thermal-history-parameters-low-z}), to define the effective equation of state of the IGM shown by the red dashed line. The final equation of state at $z=0.3$ is best described by parameters $T_0=4902 \: {\rm K}$ and $\gamma=1.53$ (black dashed line).}
{\label{fig:eos-T-15000-G-1-3}}

\subsection{Generating the Ly-$\alpha$ forest} 
\label{subsec:ly-alpha-forest}

The Ly-$\alpha$ forest is generated from the simulation box by shooting random lines of sight and storing the gas overdensities $\Delta$, component of velocities and the {\sc cite} temperatures $T$ on grid points along the line of sight .
Assuming that there is no significant evolution in the gas properties within the redshift bin, we splice together the lines of sight in such a way that it covers a redshift path identical to the observed spectra. The spectra of the Ly-$\alpha$ transmitted flux $(F)$ are generated using the procedure given in \citet{trc2001} and \citet{hamsa2015}. There are essentially four steps in simulating the spectra: (i) The temperature, baryonic density field and peculiar velocity along the sightline is calculated using Eqs. \ref{eq:smth-kernal} to \ref{eq:gadget-weigthing} as explained in \S \ref{subsec:dens-vel-temp}; (ii) The neutral hydrogen density $(n_{\rm HI})$ field along a sightline is obtained from the baryonic density field assuming photoionizing equilibrium with UVB;  (iii) The $n_{\rm HI}$ field is then used for calculating the Ly-$\alpha$ optical depth $\tau$ at each pixel accounting for peculiar velocity effects and the thermal and natural widths of the line profile; (iv) The transmitted flux is given simply by $F = \exp(-\tau)$. Note that the spectra thus generated depend on initial conditions of the model (at $z_1 = 2.1$ refer Table \ref{tab:thermal-history-parameters-low-z}) and the photoionization rate $\Gamma_{12}$ ($\Gamma_{\rm HI}$ in units of $10^{-12}$ s$^{-1}$) at the redshift of interest ($z<0.5$). To arrive at different self-consistent combinations of the two parameters $T_0$ and $\gamma$ at the redshift of interest, we vary these two parameters at the initial redshift $z_1 = 2.1$ and calculate the temperature for each gas particle using {\sc cite}\footnote{There is an apparent inconsistency in our analysis because the temperature evolution is calculated for a fixed value of the photoionization rate (e.g., that given by KS15), while we vary the same quantity $\Gamma_{12}$ at the redshift of interest treating it as a free parameter. This, however, does not affect our results as the obtained gas temperatures (at $z<0.5$) are insensitive to the assumed value of the $\Gamma_{\rm 12}$ (at $z < 2.1$).} mentioned in \S \ref{subsec:cite} and Table \ref{tab:thermal-history-parameters-low-z}.

In order to enable fair comparison with the observational data, we prepare a sample of mock spectra which has properties resembling as close as possible to the observed ones. Each redshift bin contains different number of observed spectra (see Table \ref{tab:obs-property}). Let us assume that there are $N_{\rm spec}$ observed spectra at the redshift of interest $z$. For given thermal history and a free parameter $\Gamma_{\rm HI}$, we first generate $N_{\rm spec}$ simulated spectra using the method described above, which we call a ``mock sample''. We repeat the procedure by choosing different random sightlines and generate $N$ such mock samples. We take $N=500$ in this work. The collection of $N$ mock samples constitute a ``mock suite''. Thus at $z$, the mock suite consists of $N \times N_{\rm spec}$ simulated spectra.
The velocity separation of pixels in the simulated spectra is $\sim 5 $ km s$^{-1}$ which is set by the resolution of the box, whereas the velocity resolution of observations is $\sim 17$ km s$^{-1}$. We therefore resample the simulated spectra (by linear interpolation) to match the observed data and then convolve with the line spread function (LSF) given for HST-COS spectra. HST-COS LSF\footnote{http://www.stsci.edu/hst/cos/performance/spectral \textunderscore resolution/} is given at various wavelength line centers  ($\lambda_c$) e.g. from $1150 \: {\rm \AA}$ to $1750 {\rm \AA}$ in steps of $50 {\rm \AA}$. For our purpose we assume that the broadening function is not changing over the range $\lambda_c \pm 25 {\rm \AA}$.
Finally, we add random noise to each spectrum in accordance with the SNR of the observed data, e.g., a mock sample of $N_{\rm spec}$ spectra corresponds to $N_{\rm spec}$ different values of SNR as in the observed spectra. We found that SNR varies across the spectrum in the observed data. For each spectrum we calculate the SNR in 5 different regions and choose the median SNR.  We use this observed median SNR in simulated spectra to mimic observations. For comparison, we show a simulated absorption spectrum along a random line of sight through our simulation box in the bottom panel of Fig. \ref{fig:obs-sim-spectra}. One can see that the simulated spectrum is qualitatively quite similar to the observed one (the one with removal of all other lines except Ly-$\alpha$) shown in the middle panel.

Before proceeding with the analysis of the HST-COS data within the framework of our simulated data it is important to compare our model with those in the literature. 

\subsection{Comparison with other simulations} \label{subsec:comparison-other-simulation}
We consider three predictions of our simulation that can be used for comparing different simulations.
These are (i) fraction of baryons in different phases of the $T-\Delta$ diagram, (ii) predicted IGM equation of state at $z<0.3$ and (iii) the relationship between H~{\sc i} column density ($N_{\rm HI}$) and baryon overdensity $\Delta$. Some of these predictions depend on the adopted value of $\Gamma_{\rm HI}$. For the present purpose we used $\Gamma_{\rm HI}$ consistent with QSO dominated (i.e. $f_{\rm esc} = 0$) KS15 UVB radiation model.

\subsubsection{Phase diagram of baryons}

The Ly-$\alpha$ forest is produced by relatively low density and low temperature diffuse gas. According to FGPA the mean Ly-$\alpha$ optical depth is,
\begin{equation}\label{eq:FGPA}
\tau \propto \Gamma_{\rm HI}^{-1} \: (f_d \: \Omega_b \: h^2)^2 \: \Omega_m^{-0.5}.
\end{equation}
Thus inferred $\Gamma_{\rm HI}$ from Ly-$\alpha$ will be degenerate with fraction of baryons in diffuse phase ($f_d$).
We found in the literature that different groups use different ranges in $T$ and cutoff in $\Delta$ to demarcate the phase diagram (i.e., $\Delta$ vs $T$ diagram) in 4 phases (see Fig. \ref{fig:Phase_Space_Diagram}) namely diffuse, WHIM, hot halo and condensed phase.
To make a fair comparison with other results we calculate the gas fraction in diffuse and WHIM phase as per the definitions used by the authors under consideration (see Table \ref{tab:simulation-comparison}).
\citet{smith2011} label baryons at $z = 0$ as diffuse gas if $T < 10^{5} \: {\rm K}$ and $\Delta < 1000$  and as WHIM  if $10^7 \: {\rm K} > T > 10^{5} \: {\rm K}$ and $\Delta < 1000$.
We apply the same cutoff at $z=0$ and find the diffuse and WHIM fraction to be $\sim 39.11$ per cent and $40.52$ per cent respectively which is consistent with the $\sim 40$ per cent, $40 - 50$ per cent to that of \citet{smith2011} and \citet{shull2015}.  
Note that the moderate feedback processes are included in the AMR (Adoptive Mesh Refinement) simulations of \citet{smith2011} whereas ours is a SPH simulation without any feedback.

Similarly \citet{dave2010} have incorporated momentum driven galactic outflows and various other wind models in their SPH simulations (with {\sc gadget}) which we lack. They treat baryon particles (at $z=0$) as  part of diffuse if  $T < 10^{5} \: {\rm K}$ and $\Delta < 120$ and as a part of WHIM if $T > 10^{5} \: {\rm K}$ and $\Delta < 120$  and found the fraction to be $37-43$ per cent, $23 - 33$ per cent respectively. By applying similar cutoff on $T$ and $\Delta$ at $z=0$ our diffuse and WHIM fraction turns out to be $34.09$ per cent and $28.77$ per cent respectively which is in agreement  with \citet{dave2010}.

Note that both the set of simulations discussed above have similar resolution like the one we consider here.
Unlike our simulations the simulations from the literature discussed above incorporate feedback at different levels. The close matching of baryon fraction in the diffuse phase between different models reiterate the earlier findings that the contribution of feedback effects are minor in the derived $\Gamma_{\rm HI}$ \citep{dave2010,kollmeier2014,shull2015}.

\begin{table*}
\caption{Comparison of predictions of our low-$z$ simulation with those from the literature}
\begin{threeparttable}
\centering
\begin{tabular}{ccccc}
\hline \hline
Parameters to\tnote{1}&This Work &  \citep{smith2011} &\citep{dave2010} & Analytical\tnote{2} \\ 
compare &($z=0$) &  ($z=0$) &($z=0$)  & approximation\\ \hline \hline
$T_0$\tnote{a} & 3800 - 5100 K& $\sim$ 5000 K & $\sim$ 5000 K & $\sim 2555$ K\\ 
$\gamma$\tnote{a} & 1.46 - 1.62 & $\sim$ 1.60 & $\sim$ 1.60 & $\sim 1.58$ \\  \hline
Diffuse (in per cent)\tnote{b}  & 34.09  & - & 37 - 43 & -\\ 
WHIM (in per cent)\tnote{b}  & 28.77 & - & 23 - 33 & -\\ 
Diffuse (in per cent)\tnote{c}   & 39.11 & $\sim$40 & -  & -\\
WHIM (in per cent)\tnote{c}   & 40.52 & 40 - 50  & - & - \\  \hline
$\Gamma_{\rm 12}$ & $0.12 \pm 0.03$ & $0.122$ & $\sim 0.2$&  $0.12 \pm 0.03$ \\
$\Delta_0$\tnote{d}  & 34.8 $\pm$ 5.9  & 36.9  & 38.9 & $20.6 \pm 4$ \\
$\alpha$\tnote{d}  & 0.770 $\pm$ 0.022 & 0.650 & 0.741 & $0.744 \pm 0.015$ \\  \hline \hline
\end{tabular} 
\begin{tablenotes}
			\item[a] The range in $T_0$ and $\gamma$ corresponds to different initial $T_0$ (10000 to 25000 K) and $\gamma$ (1.1 to 1.8) at $z_1 = 2.1$ see Table \ref{tab:thermal-history-parameters-low-z}. 
	        \item[b] WHIM is defined as  $T  > 10^5$ K and $\Delta < 120$ whereas the diffuse gas phase is defined as $T  < 10^5$ K and $\Delta < 120$ in \citet{dave2010}.
            \item[c] WHIM  is defined as $10^7 \: {\rm K} > T  > 10^5$ K  and $\Delta < 1000$ whereas the diffuse gas phase is defined as $T  < 10^5$ K and $\Delta < 1000$   in \citet{smith2011} \citep[also refer to][]{danforth2008}.
            \item[d] The correlation between baryon overdensity $\Delta$ and H~{\sc i} column density is expressed as $\Delta = \Delta_{0} \: N_{\rm 14}^{\alpha}$, where, $\Delta_0$ is the normalization at a fiducial H~{\sc i} column density $N_{\rm HI} = N_{\rm 14} \times 10^{14} \:  {\rm cm}^{-2}$. This relation is calculated for best fit $\Gamma_{12}$ in the redshift range $0.2 < z < 0.3$ given in bracket (Fig. \ref{fig:gamma-evolution}).
            \item[1] We notice that the \citet{paschos2009} have presented simulations for low-$z$ IGM. While they check the consistency of the mean transmitted flux for their assumed $\Gamma_{\rm HI}$, no attempt was made to measure $\Gamma_{\rm HI}$. Moreover we could not have detailed comparison with their models as the metric we use for comparison are not available for their models.
           \item[2] $T-\Delta$ relation can be obtained by equating Hubble time with net cooling time \citep{theuns1998b}. $\Delta$ vs $N_{\rm HI}$ is calculated following \citet{schaye2001} assuming Ly-$\alpha$ clouds are in hydrostatic equilibrium. 
\end{tablenotes}
\end{threeparttable}
\label{tab:simulation-comparison}
\end{table*}

\begin{table*}
\centering
\caption{Details of the thermal history considered in our simulation}
\begin{tabular}{c c c c c c c c c c c}
\hline
 &  \multicolumn{2}{c}{Initial free parameters}   & \multicolumn{8}{c}{Final parameters obtained with {\sc cite}} \\ 
 & \multicolumn{2}{c}{$z_1 = 2.1$} & \multicolumn{2}{c}{$z = 0.1$}  & \multicolumn{2}{c}{$z= 0.2$}  & \multicolumn{2}{c}{$z = 0.3$} & \multicolumn{2}{c}{$z = 0.4$}  \\ \hline 
 Model  Name & $T_0$ & $\gamma$ & $T_0$ & $\gamma$ & $T_0$ & $\gamma$ &  $T_0$ & $\gamma$  & $T_0$ & $\gamma$  \\ \hline
$T10-\gamma1.1$ & 10000 & 1.10 & 4136 & 1.54 & 4326 & 1.53 & 4589 & 1.51 & 4844 & 1.50 \\ 
$T10-\gamma1.8$ & 10000 & 1.80 & 4133 & 1.61 & 4313 & 1.61 & 4568 & 1.60 & 4810 & 1.60 \\ 
$T20 - \gamma1.1$ & 20000 & 1.10 & 4546 & 1.48 & 4971 & 1.46 & 5383 & 1.44 & 5811 & 1.42 \\ 
$T20 - \gamma1.8$ & 20000 & 1.80 & 4493 & 1.62 & 4889 & 1.62 & 5279 & 1.61 & 5677 & 1.61 \\ 
$T15 - \gamma1.3$ & 15000 & 1.30 & 4245 & 1.55 & 4583 & 1.54 & 4902 & 1.53 & 5220 & 1.51 \\ \hline
\end{tabular}
\label{tab:thermal-history-parameters-low-z}
\end{table*}

\InputFig{Phase_Space_Diagram.pdf}{80}
{Phase diagram ($T-\Delta$ plane) of randomly selected 20000 {\sc gadget-2} particle post-processed with our module {\sc cite} in our simulation at $z=0$. The black dashed line cutoff at $T=10^5$ and $\Delta = 120$ demarcates diffuse, WHIM, hot halo and condensed gas phase consistent with \citet{dave2010} (different authors use different definitions, refer to Table \ref{tab:simulation-comparison} and \S\ref{subsec:comparison-other-simulation} for details). Diffuse gas phase is mainly responsible for the H~{\sc i} absorption seen in the QSO spectrum in the form of Ly-$\alpha$ forest. The percentage of baryons in different phases are given in legend.}
{\label{fig:Phase_Space_Diagram}}

\subsubsection{Equation of state}\label{subsec:eos}
The uncertainties in the epoch of He~{\sc ii} reionization are reflected in the values of $T_0$ and $\gamma$ at the initial redshift $z_1=2.1$.
To account for this, we vary $T_0$ and $\gamma$ at $z_1=2.1$ by allowing them to take extreme values \citep[for $T_0$ and $\gamma$ measurement at high-$z$ refer][]{schaye2000,lidz2010,becker2011,boera2014} and obtain the temperatures at redshifts of our interest using {\sc cite}. The resulting values of $T_0$ and $\gamma$ at lower redshifts as obtained from the $T - \Delta$ distribution are shown in Table \ref{tab:thermal-history-parameters-low-z}. One can see that even for a widely different values of the two parameters at $z_1 = 2.1$, the equation of state at $z \sim 0.1 - 0.4$ are quite similar with $T_0 \sim 4000-6000$ K and $\gamma \sim 1.5-1.6$. This implies that the low-$z$ IGM loses, to a large extent, any memory of the He~{\sc ii} reionization. Our results are consistent with previous simulations by \citet{dave2010} and \citet{smith2011} who found that the equation of state parameters at $z=0$ are $T_0 \sim 5000$ K and $\gamma \sim 1.6$.

The equation of state at low redshifts can be derived by equating net cooling time scale with Hubble timescale.  \citet{theuns1998b} derived such relation (at high $z$) in low density regime by assuming that the heating rate is dominated by photoheating and cooling rate is dominated by recombination cooling and inverse Compton cooling. The relationship between $T$ and $T_0$ under this approximation turns out to be,
\begin{equation}\label{eq:analytical-eos}
T \sim  T_0 \: \Delta^{\frac{1}{1+\beta}}
\end{equation}
While deriving above equation, we have assumed that the H~{\sc i} recombination rate scales as $T^{-\beta}$. For $\beta = 0.7$ the slope of equation of state is $\gamma = 1.59$. This value is very much close to the one we obtained by evolving the IGM temperature using {\sc cite} thus validating our method (see Table \ref{tab:thermal-history-parameters-low-z}). 
The mean IGM temperature in the above equation at $z=0$ is $\sim 2555$ K \citep[see Eq. C21 in][]{theuns1998b}. From Table \ref{tab:thermal-history-parameters-low-z}, one can see our derived temperatures are higher by factor $\sim 2$ because  Eq. \ref{eq:analytical-eos} neglects the heating due to other sources such as shock heating, adiabatic heating due to structure formation etc.
 
The distribution of the {\sc cite} temperatures for the gas particles is shown in Fig. \ref{fig:temperature-distribution}. The left-hand panel shows the distribution at $z = 0.3$ for different initial values of $T_0$ and $\gamma$. We can see that the distributions at low-$z$ are relatively insensitive to the initial equation of state. Some small differences can be seen at lower temperatures, consistent with the equation of state given in Table \ref{tab:thermal-history-parameters-low-z}. The right-hand panel of Fig. \ref{fig:temperature-distribution} shows the {\sc cite} temperature distribution at different redshifts for model $T15-\gamma1.3$. As expected, the fraction of shock heated particles increases with decreasing redshift which is a direct consequence of structure formation shocks.

\InputFigCombine{Temperature_Distribution.pdf}{160}
{The temperature distribution of the gas particles after using {\sc cite} starting from varied initial condition at $z_1=2.1$ \emph{Left-hand panel:} the final temperature distribution at $z=0.3$ for different initial $T_0$ and $\gamma$ at $z_1=2.1$ (see Table \ref{tab:thermal-history-parameters-low-z}). \emph{Right panel:} the temperature distribution at different redshifts $z=1.5,0.9,0.3$. The initial equation of state ($z_1 = 2.1$) for right-hand panel corresponds to model $T15-\gamma1.3$ in Table \ref{tab:thermal-history-parameters-low-z}.}
{\label{fig:temperature-distribution}}

\subsubsection{$\Delta$ vs $N_{\rm HI}$ relation} \label{subsubsec:delta-NHI}
We further compared our simulations with other simulation using relation between baryon overdensity $\Delta$ and H~{\sc i} column density $N_{\rm HI}$. 
Conventionally this relation is fitted by power-law,
\begin{equation}
\label{eq:delta_b-NHI}
\begin{aligned}
\Delta = \Delta_0 \: N_{\rm 14}^{\alpha}\\
\end{aligned}
\end{equation}
where $\Delta_0$ is the normalization at fiducial H~{\sc i} column density $N_{\rm HI} = N_{14} \times 10^{14} \: {} \: {\rm cm^{-2}}$. 
Assuming hydrostatic equilibrium, \citet{schaye2001} has derived the above relation analytically for optically thin gas.
The normalization and slope of the $\Delta$ vs $N_{\rm HI}$ relation is given by,
\begin{equation}\label{eq:delta-NHI-analytical}
\begin{aligned}
\alpha &= \frac{1}{1.5-0.26 (\gamma-1)} \\
\Delta_0 &\sim \bigg[598 \: \Gamma_{\rm 12} \: {\rm T}^{0.26}_{0,4} \bigg( \frac{1.25}{1+z} \bigg)^{4.5} \bigg( \frac{0.0221}{\Omega_{\rm b} h^2} \bigg)^{1.5} \bigg( \frac{0.16}{f_g}\bigg)^{0.5} \bigg]^{\alpha}
\end{aligned}
\end{equation}
where $\gamma$ and ${\rm T}_{0,4}$ is slope of equation of state and mean IGM temperature in units of $10^4$K respectively and $f_g$ is fraction of mass in the gas (excluding stars and molecules).
For $\gamma \sim 1.6$, $T_{0,4} \sim 0.45$ and $\Gamma_{\rm 12} = 0.12 \pm 0.03$ (for $0.2 < z  < 0.3$), the slope and normalization is given as $\alpha \sim 0.744$ and $\Delta_0=20.6 \pm 4$. This simple analytic approach is known to produce $\gamma$ close to what has been seen in the simulations. However, its prediction of the normalization constant need not be accurate as one needs to take care of the baryon fraction in different phases.

To calculate such a relation in simulated spectra  we fit the Voigt profile to the absorption lines using our automatic code (see \S\ref{subsec:cdd}).
To associate the baryon overdensity with absorption line we calculate the optical depth ($\tau$) weighted overdensity $\widetilde{\Delta}$ \citep{schaye1999} as follows.
Let $\tau_{ij}$ be the optical depth contribution of overdensity $\Delta_i$ at the wavelength corresponding to the pixel $i$ to the optical depth at pixel $j$. Then the $\tau$ weighted overdensity $\widetilde{\Delta}_j$ at pixel $j$ is given by, 
\begin{equation}\label{eq:optical-depth-weighted-delta}
\begin{aligned}
\widetilde{\Delta}_j &= \frac{\sum \limits_{i=1}^{N} \tau_{ij} \Delta_i}{\sum \limits_{i=1}^{N} \tau_{ij}},
\end{aligned}
\end{equation}
where $N$ is the total number of pixels in the spectrum. The total optical depth at a pixel $j$ is given by,
\begin{equation}\label{eq:optical-depth-sum}
\begin{aligned}
\tau_j &= \sum \limits_{i=1}^{N} \tau_{ij}.
\end{aligned}
\end{equation}
Fig. \ref{fig:optical-depth-weighted-delta} demonstrates our procedure for assigning $\tau$ weighted overdensity to an absorption line.
1$^{\rm st}$,2$^{\rm nd}$,3$^{\rm rd}$ and 4$^{\rm th}$ panel from top shows flux, overdensity $(\Delta)$, temperature $(T)$ and peculiar velocity $(v)$ respectively.
The flux in the top panel is calculated from $\Delta$, $T$ and $v$ (all solid blue lines).
As expected due to power-law equation of state, $\Delta$ and $T$ are correlated (solid blue lines).
We calculate the $\tau$ weighted temperature by replacing $\Delta_i$ in Eq. \ref{eq:delta-NHI-analytical} by $T_i$.
The $\tau$ weighted overdensity and the $\tau$ weighted temperature (shown by red dashed line in 2$^{\rm nd}$ and 3$^{\rm rd}$ panel from top) are also correlated.
We then associate this $\tau$ weighted overdensity at the absorption line center to the column density of that line (obtained by Voigt profile fitting). 
Fig. \ref{fig:delta_NHI_scatter} shows the density plot of $\tau$ weighted overdensity $\Delta$  and column density $N_{\rm HI}$ for 4000 simulated Ly-$\alpha$ spectra (SNR = 50) in the range $0.2<z<0.3$ for $\Gamma_{\rm 12} = 0.12$.
The magenta errorbar shows the mean $\tau$ weighted overdensity (with $1\sigma$ error) in each bin of size $\Delta \log{N_{\rm HI}=0.1}$.
The black dashed line shows the power-law fit with $\Delta_0 = 34.8 \pm 5.9$ and $\alpha = 0.77 \pm 0.022$.  
The error in $\Delta_0$ corresponds to 1$\sigma$ range in $\Gamma_{\rm 12}$ (see \S\ref{subsec:Gamma12-evolution}).
We find that $\alpha$ is less sensitive to $\Gamma_{\rm 12}$, the error in $\alpha$ accounts for different thermal history (i.e., different values of $\gamma$).

\InputFig{tau_weighted_3_new}{85}{Illustration of assigning optical depth ($\tau$) weighted overdensity and temperature to the absorption lines in simulated spectrum (see \S\ref{subsubsec:delta-NHI}). The flux shown in the \emph{top} panel is computed from overdensity $(\Delta)$ (blue solid line), temperature $(T)$ (blue solid line) and peculiar velocity $(v)$ (blue solid line) given in 2$^{\rm nd}$,3$^{\rm rd}$ and 4$^{\rm th}$ panel from top respectively. The $\tau$ weighted overdensity (see Eq. \ref{eq:optical-depth-weighted-delta}) and $\tau$ weighted temperature are shown by red dashed lines in 2$^{\rm nd}$ and 3$^{\rm rd}$  panel from top respectively. To calculate $\tau$ weighted temperature, we replaced $\Delta_i$ in Eq. \ref{eq:optical-depth-weighted-delta} by $T_i$.}
{\label{fig:optical-depth-weighted-delta}}  

\InputFig{Delta_vs_NHI_image.pdf}{80}
{Correlation between $\tau$ weighted overdensity $\Delta$ (see \S\ref{subsubsec:delta-NHI}) and  column density $N_{\rm HI}$ for $4000$ simulated Ly-$\alpha$ forest spectra (SNR=50) in the range $0.2 < z < 0.3$ for $\Gamma_{\rm 12} = 0.12$ (consistent with our final measurements see Fig. \ref{fig:gamma-evolution}). The color scheme represents the density of points in logarithmic units.  Magenta star points with errorbar are mean $\tau$ weighted overdensity binned in $N_{\rm HI}$ with width $\Delta {\rm  N}_{\rm HI} = 0.1$. Black dashed line shows our best fit to the mean $\tau$ weighted overdensity. The errorbar on best fit values corresponds $1\sigma$ variation in $\Gamma_{\rm 12}$ which is 0.03 (see Fig. \ref{fig:gamma-evolution}).} 
{\label{fig:delta_NHI_scatter}}

The power-law index from our simulation ($\alpha = 0.77 \pm 0.022$) matches well (within $2\sigma$) with \citet{dave2010} and \citet{tepper2012}($\alpha \sim 0.786$) and theoretically expected values but is slightly higher than \citet{shull2015} (see Table \ref{tab:simulation-comparison}).
The normalization parameter $\Delta_0$ in our simulation $34.8 \pm 5.9$ is in agreement (within $1\sigma$) with \citet{smith2011,shull2015} and \citet{dave2010} simulation but less than \citet{tepper2012} ($\Delta_0 \sim 48.9$).
Note that \citet{tepper2012} used a simulation box with factor $2$ lower resolution.
Also $\Delta_0$ is sensitive to the value of $\Gamma_{\rm 12}$ used to generate simulated spectra (see Eq. \ref{eq:delta-NHI-analytical}).

We also found that the mean flux decrement (DA) calculated from our simulation is within $30$ percent to that found from \citet{paschos2009} and within $20$ percent to that from \citet{kollmeier2014,viel2016} when we use their $\Gamma_{\rm 12}$ in our simulations. The higher values of \citet{paschos2009} can be attributed to the additional heating incorporated in their simulations. Whereas the differences in DA with \citet{kollmeier2014,viel2016} can be attributed to the differences in fraction of baryons in diffuse phase, mean IGM temperature of the gas in simulation and cosmological parameters used.

We summarize the comparison between our simulation results and others in Table \ref{tab:simulation-comparison}. As can be seen, our estimates of $T_0$ and $\gamma$, the fraction of baryons in diffuse gas and in WHIM (accounting for differences in the precise definition), as well as $\Delta$ vs $N_{\rm HI}$ relation match quite well with those from  other simulations, thus validating our method of evolving the gas temperature using {\sc cite}. The advantages of using {\sc cite} for the low-$z$ IGM studies are as follows:
\begin{itemize}
\item Because {\sc cite} is based on post-processing the {\sc gadget-2} output, the method is computationally less expensive.
\item {\sc cite} allows us to explore large thermal history parameter space without performing the full SPH simulation from high-$z$.
\item We are able to include the ionization and thermal evolution in {\sc gadget-2} as post processing steps.
\item It is easy to incorporate the radiative cooling for a wide range of metallicities in {\sc cite}.
\end{itemize}
All these allow us to explore a wide range of parameter space thereby enable us to have a reliable estimation of errors associated with the derived $\Gamma_{\rm HI}$.  \\

\subsubsection{Feedback Processes}
Our simulations does not include any feedback processes such as galactic winds or AGN feedback. Previously \citet{shull2015} used the simulations of \citet{smith2011} which included two types of feedback namely \emph{local} and \emph{distributed} feedback. 
They found that the phase diagram converges as long as simulation box size $L \geq 50h^{-1}$ cMpc.
While the fraction of baryons in WHIM and condensed phase changes considerably with feedback prescription, the fraction of baryons in diffuse phase that is responsible for Ly-$\alpha$ forest remains similar.  
Because of this they found that both feedback processes affect column density distribution mildly and hence $\Gamma_{\rm 12}$ constraint remains similar \citep[see Fig. 6 in][]{shull2015}. The feedback method however, can affect the clumpiness of the IGM on small scales whereas large scale correlation between parameters such as $\Delta$ and $N_{\rm HI}$ are nearly unaffected. Similarly \citet{dave2010} and \citet{kollmeier2014} found that the properties of Ly-$\alpha$ forest are largely insensitive to wind models and feedback prescriptions. 
Keeping this in mind, now we shall use our simulation to derive $\Gamma_{\rm HI}$ from the HST-COS data.

Despite offering flexibility the obvious shortcoming of {\sc cite} is that the diffuse gas is evolved dynamically at effectively zero pressure (because of its low temperature), rather than the pressure it would have if it were at $T \sim 10^4$ K typical of photoionized gas. Thus dynamical impact of diffuse IGM pressure is not modeled self-consistently in {\sc cite}. However, the comparisons of {\sc cite} with other simulations discussed above suggests that this is not severe shortcoming. The consistency of the phase distribution and the equation of state with other simulations seem almost guaranteed by the nature of {\sc cite}. However, the consistency of $\Delta$ vs ${\rm N_{HI}}$ is non-trivial and suggests that the evolving hydrodynamic simulation with low pressure is not distorting the properties of Ly-$\alpha$ forest for the spatial resolution typically achieved in the low-z simulations that are used to reproduce HST-COS data.

\section{Flux statistics}
\label{sec:method}

In order to carry out comparison between simulations and observed data at each redshift bin given in Table \ref{tab:obs-property}, we consider two statistics of the Ly-$\alpha$ transmitted flux, namely the flux PDF and the flux PS.
In the following subsections we describe the method of calculating the flux PDF and PS and appropriate covariance matrix from the simulation.

\subsection{Flux PDF} 
\label{subsec:flux-pdf}

We compute the flux PDF of the observed and simulated spectra for all four redshift bins given in Table \ref{tab:obs-property}.
We evaluate the distribution using 11 flux bins of width $\Delta F = 0.1$ with the first bin center at $F = 0$ and last one at $F = 1$. The pixels with $F > 1$ (respectively, $F < 0$) are included in the last (respectively, first) bin.
Note that the flux bin widths used in this work are larger than those used previously at high-$z$.
This is mainly because the SNR in the present sample is much lower than what is typically achieved in high-$z$ echelle spectra \citep{jenkins1991,mcdonald2000,kim2007,vincent2007,bolton2008,rollinde}.
This will influence flux PDF in the bins near continuum and possibly introduce a strong correlation between different bins if bin width is small.

Any meaningful statistical comparison requires errors on the observed flux PDF at each flux bin, along with the noise covariance between different bins. One standard way of estimating the error covariance matrix from the observed data is to use the jack-knife method. However, we found that the errors obtained using this method are considerably underestimated. This is possibly because the cosmic variance is not properly accounted for in the jack-knife method \citep{rollinde}. Hence, we use the simulated mock samples to compute the covariance matrix as explained below:

As discussed earlier, for each redshift bin (see Table \ref{tab:obs-property}) we generate $N=500$ simulated mock samples for the free parameter $\Gamma_{\rm 12}$ and for each model given in Table \ref{tab:thermal-history-parameters-low-z}. We remind the reader that each mock sample consists of number of sightlines equal to observed number of sightlines  in the corresponding redshift bin. Let $P_{n}(F_i)$ denote the value of the flux PDF in the $i^{th}$ bin of $n^{th}$ mock sample, where $n$ takes values from 1 to $N$. Let the flux PDF in the $i^{th}$ bin averaged over all mock samples be denoted as $\overline{P}_i$ .  
The covariance matrix element $C(i,j)$ between the $i^{th}$ and $j^{th}$ bins is given by,
\begin{equation}
\label{eq:covar-terms}
\begin{aligned}
C(i,j) &= \frac{1}{N-1} \sum \limits_{n=1}^{N} \; [P_{n}(F_i) - \overline{P}_i] \; [P_{n}(F_j) - \overline{P}_j]
\end{aligned}
\end{equation}
where, $i$ and $j$ can take values from 1 to the number of bins (which in this case is 11). The covariance matrix $C$ is calculated for free parameter $\Gamma_{\rm 12}$ and for each initial $T_0-\gamma$ (at $z_1=2.1$) model given in Table \ref{tab:thermal-history-parameters-low-z}. 

To visualize the covariance matrix we calculate the correlation matrix defined as
 \begin{equation}
\label{eq:corr-terms}
\begin{aligned}
Corr(i,j) &= \frac{C(i,j)}{\sqrt{C(i,i) \: C(j,j)}}.
\end{aligned}
\end{equation}
The left-hand panel of Fig. \ref{fig:corr-matrix} shows the correlation matrix for the flux PDF for simulated Ly-$\alpha$ forest at $z=0.3$ for a model $T15-\gamma1.3$ (refer Table \ref{tab:thermal-history-parameters-low-z}) and $\Gamma_{\rm 12} = 0.1$. It is clear from the figure that the off-diagonal terms of the matrix are not negligible, thus showing that the errors in different bins are correlated. This implies that the full covariance matrix should be used to compute the $\chi^2$ while comparing the simulation with the data. We find that the correlation is strongest for the immediately neighboring bins. Also, the correlation between the neighboring bins increases for higher flux bins since large number of pixels are in the continuum. 

\InputFigCombine{Corr.pdf}{140}{Correlation matrix for the flux PDF (\emph{left-hand panel}) and the flux PS (\emph{right-hand panel}). Both the correlation matrices are calculated using the covariance obtained from the simulated mock samples (see \S\ref{subsec:flux-pdf} and \S\ref{subsec:flux-ps} for more details). The correlation matrices are shown for simulated Ly-$\alpha$ forest at $z=0.3$, with $\Gamma_{\rm 12} = 0.1$ for $T15-\gamma1.3$ model given in Table \ref{tab:thermal-history-parameters-low-z}.}
{\label{fig:corr-matrix}}  

In the observed spectra a large number of pixels are found to be in the continuum. Consequently,  the errorbar on the flux PDF in the flux bins close to continuum $F \geq 0.9$ are very small. Any $\chi^2$ minimization procedure thus tries to give more weight to the flux PDF bins around $F \geq 0.9$. In addition, these bins near continuum are affected by noise (where, SNR varies from $5$ to $15$) and continuum fitting uncertainty. This can introduce an additional correlation between bins near the continuum.
To avoid such difficulties, during $\chi^2$ minimization we used the flux PDF in the range $0 \leq F \leq 0.8$. However to normalize flux PDF we used all the bins. Note that \citet{rollinde} have used a similar cutoff to calculate the flux PDF at $z \sim 2 - 3$.  In our case it is not only the continuum uncertainty but also the relatively poorer SNR of the observed spectra, which affects the flux PDF calculation for bins with $F \geq 0.9$, are important. We have checked and found that ignoring the points near the continuum does not affect our constraints on $\Gamma_{12}$, except for a marginal increase in the errorbars on $\Gamma_{\rm 12}$.

The $\chi^2$, which will be used for quantifying the match between the observed flux PDF and with the simulated one, can be written in the matrix form as,
\begin{equation}
\label{eq:chi-sq-calculation}
\begin{aligned}
\chi^2_{(T_0,\gamma,\Gamma_{12})} = [P_{(T_0,\gamma,\Gamma_{12})} - P_{obs}] \; C^{-1} \; [P_{(T_0,\gamma,\Gamma_{12})} - P_{obs}]^{T},
\end{aligned}
\end{equation}
where $P_{obs}$ denotes observed flux PDF obtained from all the spectra in the relevant redshift bin and $P_{(T_0,\gamma,\Gamma_{12})}$ is the flux PDF obtained from our simulations for a particular model given in Table \ref{tab:thermal-history-parameters-low-z} and a free parameter $\Gamma_{\rm 12}$. Note that both $P_{obs}$ and $P_{(T_0,\gamma,\Gamma_{12})}$ are row vectors with their $i^{th}$ element being the flux PDF in the $i^{th}$ bin.
We re-emphasize that the covariance matrix $C$ is calculated in each redshift bin for each model in Table \ref{tab:thermal-history-parameters-low-z} and $\Gamma_{\rm 12}$. 
Note that, because the flux PDF is normalized, the covariance matrix is singular. Hence we use the singular value decomposition method \citep{press1992} to compute the $\chi^2$. The $\chi^2_{(T_0,\gamma,\Gamma_{12})}$ can be calculated for each combination of the free parameters $T_0 - \gamma$ at an initial redshift and $\Gamma_{\rm 12}$ at the redshift of our interest. The best-fit parameters are obtained by finding the location of the minimum of the $\chi^2_{(T_0,\gamma,\Gamma_{12})}$.
The $1\sigma$ confidence level corresponds to the region between $\chi^2_{1 \sigma} = \chi^2_{\rm min} \pm \Delta \chi^2_{1 \sigma}$, where $\chi^2_{\rm min}$ is the minimum value of the $\chi^2$ and $\Delta \chi^2_{1 \sigma} = 1$ \citep{press1992}.

\subsection{Flux PS} 
\label{subsec:flux-ps}

We compute the flux PS from observational data and simulations in redshift bins same as those used for estimating the flux PDF. However, there is a crucial difference in how we treat the sightlines while calculating the flux PS from that for the flux PDF. In the case of the flux PDF, we spliced different sightlines from the simulation box to construct a redshift path length as large as the observed redshift range. However, such splicing may introduce spurious effects while computing the two point correlation properties of the flux. Hence for calculating the flux PS we use sightlines of comoving length equal to the simulation box size $50 \: h^{-1}$ cMpc. In order to ensure that the simulations and the observations are treated on equal footing, we divide the observed spectra within each $z$ bin into segments which have comoving length equivalent to $50 \: h^{-1}$ cMpc.
To calculate the flux PS, we first compute the Fourier transform $F(k)$ of the transmitted flux. The corresponding power is given by $P(k) \propto |F(k)|^2$, where the normalisation is calculated from the condition
\begin{equation}
\label{eq:ps-normalization}
\begin{aligned}
\sigma_F^2 &= \int \limits_{-\infty}^{\infty} \frac{dk}{2\pi} \: P(k)
\end{aligned}
\end{equation}
with $\sigma_F^2$ being the variance of the transmitted flux. 
Once we compute $P(k)$ for each segment (of comoving length  $50 \: h^{-1}$ cMpc) of the spectra, we take an average over all the segments to get an estimate for the flux PS.

A reliable estimate of the flux PS can only be obtained in a limited range of scales because of the finite length of the spectra and other systematic effects. For the observed spectra, the small scale power is affected by the presence of the narrow metal lines \citep{prats2015}. Following \citet{mcdonald2000}, we choose to work with scales corresponding to $k < k_c \sim 8 \;h \; {\rm cMpc}^{-1}$. Similarly, the large scale power in the observed spectra is affected by the uncertainties in the continuum fitting \citep{kim2004}. In addition, while computing the flux PS from our simulations, the finite periodic box size of $50 \: h^{-1}$ cMpc implies that scales with $k < k_t \sim 0.2 h  \: {\rm cMpc}^{-1}$ may not be sampled properly. Hence, to make any meaningful comparison between the observations and simulations we restrict the flux PS measurements to scales $0.209 \leq k / (h~{\rm cMpc}^{-1}) \leq 8$.

Care must be exercised while binning the $P(k)$ over different $k$-ranges. We find that for $k \gtrsim 1 h \: {\rm cMpc}^{-1}$, the $k$ modes are sampled densely enough to divide the $P(k)$ into bins. In that case we use logarithmic bins of size $\log_{10}(1.24)$ similar to \citet{kim2004}. On the other hand, for $k \lesssim 1 h \: {\rm cMpc}^{-1}$, the $k$ modes are sparsely sampled and hence we do not bin the data as done in \citet{prats2015}. We ensure that the same procedure is followed while dealing with the observed and simulated spectra.

The procedure for computing the errorbars on the observed flux PS, i.e., estimating the error covariance matrix $C$, is same as the one discussed for the flux PDF. The matrix $C$ is estimated from the suite of mock spectra. We also attempted to estimate it from the observed spectra using the jack-knife method. However, since the number of observed spectra is relatively small, the off-diagonal terms in the covarinace matrix are found to be noisy. The right-hand panel of Fig. \ref{fig:corr-matrix} shows the flux PS correlation matrix. 
The flux PS correlation matrix is dominated by diagonal terms consistent with \citet{mcdonald2000} and \citet{zhan2005}. Since the neighboring $k$-modes are likely to be correlated, the neighboring bins show strong correlation.  As one moves away from the diagonal terms the correlations between the different mode decrease. The smaller scales show slightly stronger correlations because the underlying density field is more non-linear and non-gaussian at these scales \citep{zhan2005}.

We follow the same procedure as discussed in the previous section for calculating $\chi^2$ between model and observed flux PS.

\subsection{Tests with the mock spectra} 
\label{subsec:test-with-mock-spectra}

Before using the observed and simulated spectra, to constrain the photoionization rate $\Gamma_{12}$ at $z<0.45$, we carry out a few tests on the simulated quantities. First, we check the sensitivity of the flux PDF and flux PS on the parameters $T_0 - \gamma$ (at initial redshift $z_1=2.1$) and $\Gamma_{12}$. Fig. \ref{fig:flux-pdf-ps-variation} shows the dependence of the flux PDF (left-hand panel) and flux PS (right-hand panel) at $z = 0.3$ on these three parameters. Note that the values of $T_0$ and $\gamma$ at the redshift of interest are obtained by varying these two parameters at the initial redshift $z_1 = 2.1$.
The solid and dashed lines, for a given combination of $T_0$ and $\gamma$, in the Fig. \ref{fig:flux-pdf-ps-variation} correspond to $\Gamma_{\rm 12} = 0.08$ and $0.12$ respectively at $z \sim 0.3$.

Even though we varied $T_0$ at $z_1=2.1$ by factor of 2, the value of $T_0$ obtained at $z \sim 0.3$ using {\sc cite} differ by only $\sim 9$ per cent. Similarly, for given $T_0$ at $z_1=2.1$ even when we change initial $\gamma$ between $1.1$ and $1.8$ values of $T_0$ and $\gamma$ obtained at $z \sim 0.3$ are nearly identical. Thus the flux PDF and PS are fairly insensitive to our choice of $T_0 - \gamma$ at $z_1=2.1$. In other words the flux PDF is insensitive to the He~{\sc ii} reionization history (also see \S\ref{subsec:eos}). It is also clear from Fig. \ref{fig:flux-pdf-ps-variation} that both the above statistics are sensitive to assumed value $\Gamma_{\rm 12}$. Therefore they can be used to constrain $\Gamma_{\rm 12}$.

We next study how well the two statistics can be used for constraining the $\Gamma_{12}$. For a  particular model given in Table \ref{tab:thermal-history-parameters-low-z} and a fixed value of $\Gamma_{\rm 12}$ (at $z=0.3$), we construct a mock sample from the simulations, i.e., a set of sightlines which have properties similar to the observed ones. This mock sample can be treated as the ``input'' data for which the two statistics (flux PDF and PS) can be calculated. We then draw sightlines from other simulation box (parameters $T_0$ and $\gamma$ are different from that of the input data) to construct a large number $(N = 500)$ of mock samples and compute the two statistics along with the error covariance matrix. The input data and mock samples are drawn from two different simulation boxes with identical cosmological parameters but different initial conditions. The idea is to vary the value of $\Gamma_{12}$ for these samples and compare with the input data. The minimization of the $\chi^2$ should enable us to obtain the best-fit value of $\Gamma_{12}$ along with the errorbars, which can be compared with the input value of $\Gamma_{12}$. The result of the analysis for seven different input $\Gamma_{12}$ values for $z \sim 0.3$ is shown in Fig. \ref{fig:recovery}. The red dashed line indicates the case where there is perfect match between the input and the recovered $\Gamma_{\rm 12}$. The point with errorbars show the recovered $\Gamma_{12}$ with the 1$\sigma$ confidence interval for each input model.  We can see that our analysis recovers the input value of the photoionization rate quite accurately. The typical statistical uncertainty in recovering $\Gamma_{\rm 12}$ is  $\sim 0.015$.

\InputFigCombine{Flux_PDF_PS_Variation.pdf}{160}
{\emph{Left} and \emph{right-hand panels} show respectively the variation of the flux PDF and PS (at $z = 0.3$) for different models given in Table \ref{tab:thermal-history-parameters-low-z} and  $\Gamma_{\rm 12} = 0.08$ and $0.12$ at $z=0.3$.  The values of $\Gamma_{\rm 12}$ and model corresponding to different lines are indicated in the legend. It is clear from the figure that flux PDF and PS are more sensitive to $\Gamma_{\rm 12}$ (at $z=0.3$) than initial values of $T_0$ and $\gamma$ (at $z_1=2.1$) because the final equation of states at $z=0.3$ are very similar (see Table \ref{tab:thermal-history-parameters-low-z}).}
{\label{fig:flux-pdf-ps-variation}}

\InputFig{Recovery.pdf}{80}
{Recovery of the $\Gamma_{\rm 12}$ at $z = 0.3$ using the flux PDF and PS and $\chi^2$ statistics. The $x$-axis represents the true $\Gamma_{12}$, i.e., the one used in the input model. The points with errorbars show the recovered $\Gamma_{12}$ with the 1$\sigma$ confidence interval for each input model. The red dashed line indicates the case where there is perfect match between the input and the recovered $\Gamma_{\rm 12}$. The input and mock data are drawn from two different simulations with same cosmological parameters but different initial condition. The typical uncertainty in recovered $\Gamma_{\rm 12}$ is $\pm 0.015$.}
{\label{fig:recovery}}

\InputFigCombine{Instr_Broad.pdf}{160}
{\emph{Left-hand} and \emph{right-hand panels} show the effect of LSF on flux PDF and PS respectively. In both panels results obtained with Gaussian LSF (FWHM $\sim 17$ km s$^{-1}$) are shown using solid blue curves and ones that are obtained using HST-COS LSF are shown by red dashed curves. In left-hand panel number of saturated pixels (i.e. $F \sim 0$) are smaller when we use the HST-COS LSF. Right-hand panel shows that the LSF affects the overall normalization ($\sigma^2_{F}$) of the flux PS below $k \sim 6 \: h \: {\rm cMpc}^{-1}$.}
{\label{fig:instrumental-broadening-effect}}

Finally, we test the effect of using the HST-COS LSF instead of the traditionally used Gaussian profile function. Fig. \ref{fig:instrumental-broadening-effect} shows the flux PDF (\emph{left-hand panel}) and the flux PS (\emph{right-hand panel}) obtained using the two LSFs. The Gaussian LSF used for making this plot has an FWHM $= 17$ km s$^{-1}$ as shown by solid blue curve. The HST-COS LSF is asymmetric and has extended wings that do not go to zero as rapidly as Gaussian LSF. Hence, the number of pixels near zero are less in HST-COS LSF as compared to Gaussian LSF (\emph{left-hand panel}). The LSF also affects the variance of flux field ($\sigma^2_F$) and hence the normalization of flux PS (\emph{right-hand panel}).  We found that one would overpredict the $\Gamma_{\rm 12}$ by $\sim 20$ per cent if the Gaussian LSF is used in the simulated spectra instead of HST-COS LSF using flux PS.

\section{Results and Discussion}
\label{sec:results}

\subsection{Constraints on $\Gamma_{\rm 12}$} \label{subsec:Gamma12-constraint}

\begin{figure*}
\centering
\graphicspath{{./Images/}}
\includegraphics[width=160mm]{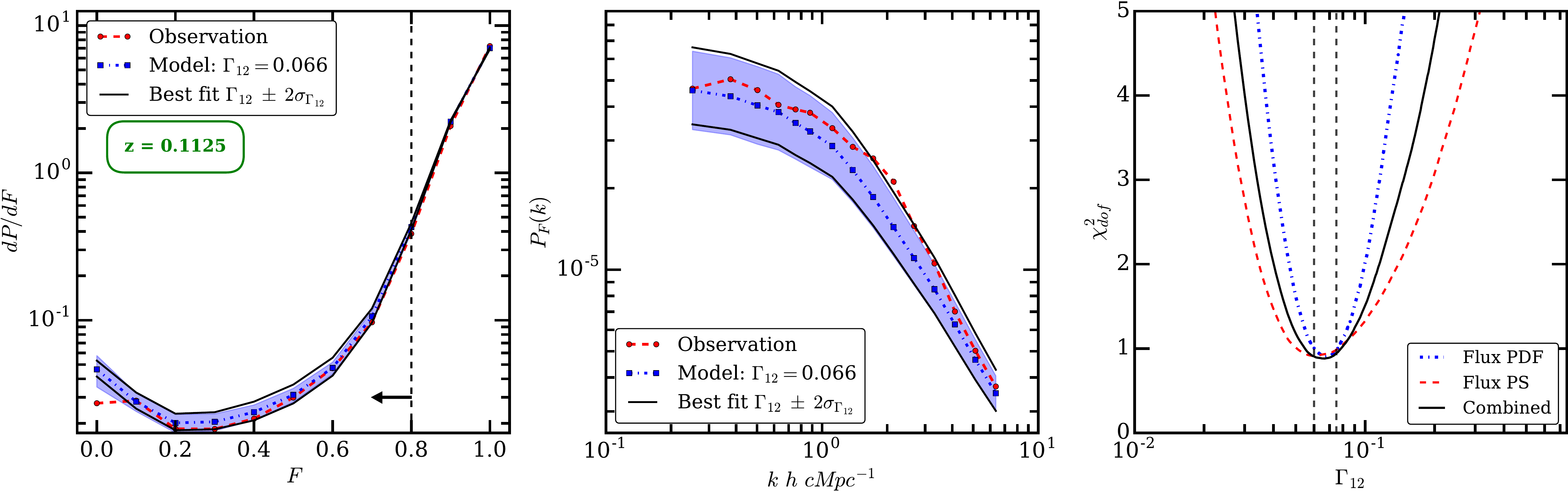}
\includegraphics[width=160mm]{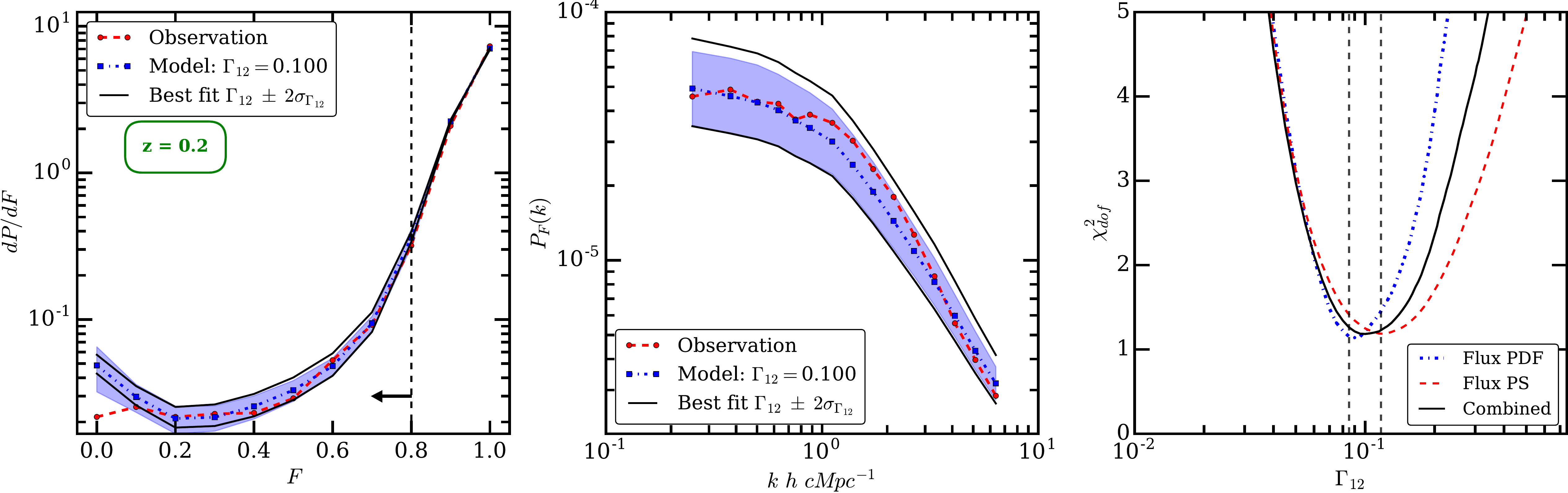}
\includegraphics[width=160mm]{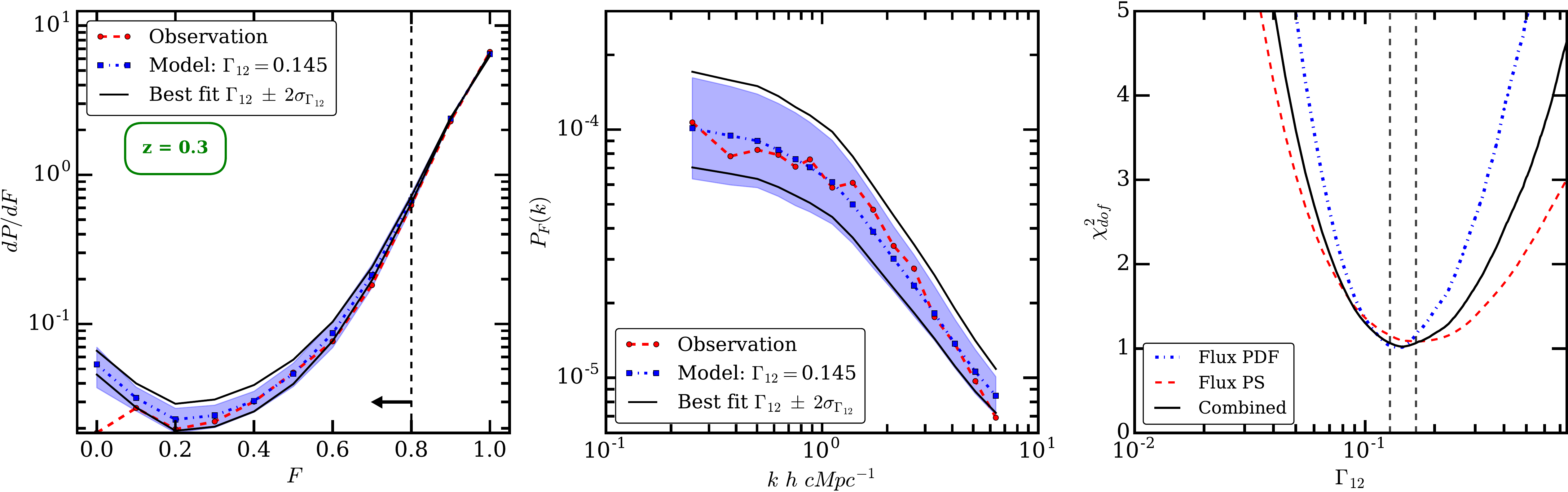}
\includegraphics[width=160mm]{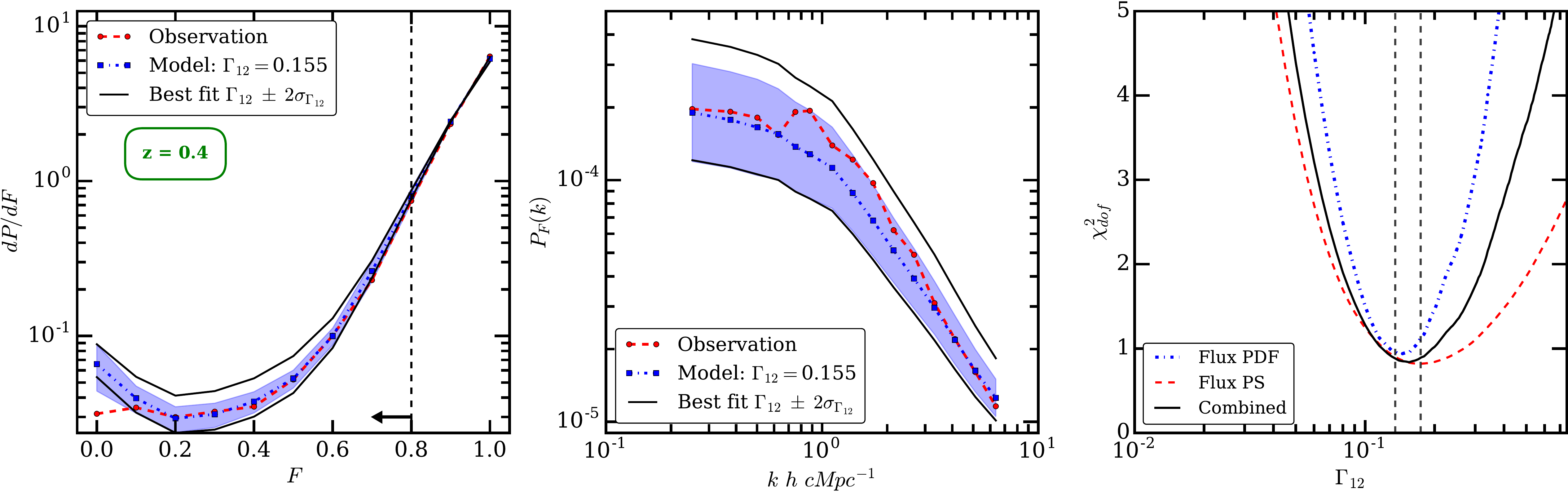}
\caption{Constraints on $\Gamma_{\rm 12}$ for $z = 0.1125,0.2,0.3,0.4$ from top row to bottom row. The left-hand panels show the flux PDF of observed (red dashed line) and best fit model spectra (blue dotted line). The blue shaded regions show the $1\sigma$ range in flux PDF from mock samples covariance matrix (diagonal terms only). 
The middle panels  show the flux PS of observed (red dashed line) and best fit model spectra (blue shaded region). The blue shaded regions show the $1\sigma$ range in flux PS from mock samples  covariance matrix (diagonal terms only). 
The right-hand panels show the reduced $\chi^2$ against $\Gamma_{\rm 12}$ for flux PDF (blue dotted curve),  flux PS (red dashed curve) and the combined statistics (i.e., flux PDF and PS). The black solid curve is obtained by adding the $\chi^2$ of the flux PDF and PS. Note that the best fit model corresponds to minimum value of joint $\chi^2_{dof}$. The vertical black dashed lines show the $1\sigma$ (statistical only) constraint on $\Gamma_{\rm 12}$. For visual purpose, the model flux PDF and PS obtained by shifting the best fit $\Gamma_{\rm 12}$ by $\pm 2 \sigma$ range is shown (black solid line) in the left-hand and middle panel respectively. All the plots are shown for model $T20-\gamma1.8$ (see Table \ref{tab:thermal-history-parameters-low-z}). For $z=0.4$ redshift bin results are shown for the simulated spectra that are \emph{not} contaminated by Ly-$\beta$ forest (see text and Fig. \ref{fig:ly-beta-contamination} for details).}
\label{fig:result-hst-cos-2015-2}
\end{figure*}

We obtain constraints on $\Gamma_{12}$ by comparing the flux PDF and PS from the simulated Ly-$\alpha$ forest with those from the HST-COS data using the $\chi^2$-minimization technique discussed in \S\ref{sec:method}. Fig. \ref{fig:result-hst-cos-2015-2} shows the results of our analysis for the four redshift bins identified in Fig. \ref{fig:redshift-coverage} and given in Table \ref{tab:obs-property} i.e., for $z=0.1125,0.2,0.3$ and $0.4$. $T_0$ and $\gamma$ at these redshifts are obtained for different model (see Table \ref{tab:thermal-history-parameters-low-z}) by  using {\sc cite}. For a particular model there is one free parameter $\Gamma_{\rm 12}$ at each $z$.

The left-hand panels in the Fig. \ref{fig:result-hst-cos-2015-2} (all results are shown for model $T20-\gamma1.8$) show the comparison of observed flux PDF (red dashed curve) with that of the best fit model (blue dotted curve). The blue shaded regions indicate the 1$\sigma$ dispersion (i.e., corresponding to diagonal term of the covariance matrix in Eq. \ref{eq:covar-terms}) in model flux PDF at each value of $F$ calculated from the mock sample. Although we plot the flux PDF in the range $0$ to $1$, we use only the flux bins $0 \leq F \leq 0.8$ for the $\chi^2$ analysis (as indicated by black dashed vertical line with arrow). We find that the match between the simulated spectra and the observed ones are quite good for all the four redshifts (typical $\chi^2$ per degree of freedom i.e., $\chi^2_{\rm dof} \sim 1$). The only bin where the two do not agree is the bin with $F = 0$ where the observed spectra systematically predict less number of pixels. This mismatch could be due to uncertainty in background subtraction in HST-COS data.

The middle panels show the comparison of observed flux PS (red dashed curve) with best fit model (blue dotted curve). The blue shaded regions indicate the 1$\sigma$ range on the model flux PS. It is interesting to see that the match again is quite good, and we can match both the flux PDF and PS for the same value of $\Gamma_{12}$.

The right-hand panels show the variation of the reduced $\chi^2$ with $\Gamma_{\rm 12}$ for the flux PDF (blue dotted curve), the flux PS (red dashed curve) and the combined (i.e., flux PDF and PS) statistics (black solid curve). The best fit $\Gamma_{\rm 12}$ used in the left-hand and middle panels corresponds to the one which gives minimum $\chi^2$ for the combined case. We see that all the three curves have the expected parabolic shape and the reduced $\chi^2$ (i.e., $\chi^2_{\rm dof}$) at the minima have values $\sim 1$. The minimum of the $\chi^2$ for both the flux PDF and PS occur at similar values of $\Gamma_{12}$ and agree well within 1$\sigma$. We also note that the width of the curve is smaller for the flux PDF alone case than for the flux PS alone case which implies that the flux PDF is better at constraining $\Gamma_{\rm HI}$ than the flux PS. However, flux PDF is sensitive to continuum fitting uncertainty in the HST-COS data,  whereas flux PS is less sensitive to it as we will discuss later. 

The errors on $\Gamma_{12}$ obtained using the above method would be an underestimation since it ignores various other effects. We account for possible statistical and systematics uncertainties in $\Gamma_{\rm 12}$ in Table \ref{tab:error-budget}. The first of these is the uncertainty in the thermal history, i.e., the two initial  parameters $T_0$ and $\gamma$. Ideally one should vary the three parameters simultaneously, obtain the joint likelihood and marginalize over all other parameters except $\Gamma_{12}$. However, since we need to solve the temperature evolution for each $T_0- \gamma$ combination, the full analysis can be quite computationally expensive. Hence we take a slightly different approach where we choose the most extreme values of $T_0$ and $\gamma$ at $z_1=2.1$ compared to observations \citet{schaye2000,becker2011} and \citet{boera2014} \citep[for compilation see][]{puchwein2015}. For each combination of $T_0$ and $\gamma$ (given in Table \ref{tab:thermal-history-parameters-low-z}), we evolve the thermal history to lower redshifts, compute the Ly-$\alpha$ forest and obtain $\Gamma_{12}$. The constraints for different thermal histories for the four redshift bins are summarized in Tables  \ref{tab:Gamma-constraint-1}, \ref{tab:Gamma-constraint-2}, \ref{tab:Gamma-constraint-3} and \ref{tab:Gamma-constraint-4}. It is clear from the tables that the constraints on $\Gamma_{12}$ are relatively insensitive to the values of $T_0$ and $\gamma$ at $z_1 = 2.1$. This is what we have seen in Fig. \ref{fig:flux-pdf-ps-variation}.
In all cases the differences are small and within statistical uncertainty so for quoting best fit $\Gamma_{\rm 12}$, we used model $T20-\gamma1.8$ which has minimum $\chi^2$ in three out of four redshift bins. 

\begin{table}
\caption{Joint (Flux PDF + Flux PS) $1\sigma$ constraint on $\Gamma_{\rm 12}$ for different thermal history (see Table \ref{tab:thermal-history-parameters-low-z}) for redshift bin I (refer Table \ref{tab:obs-property})}
\begin{threeparttable}
\centering
\begin{tabular}{ccccc}
\hline
 & \multicolumn{2}{c}{$z=0.1$}  & \multicolumn{2}{c}{Joint constraint} \\ \hline
Model & $T_0$ & $\gamma$ & $\Gamma_{\rm 12}$\tnote{a} & $\chi^2_{\rm dof}$ \\ \hline
$T10-\gamma1.1$ & 4136 & 1.54 & 0.067 $\pm$ 0.008 & 0.83 \\ 
$T10-\gamma1.8$ & 4133 & 1.61 & 0.068 $\pm$ 0.007 & 0.83 \\ 
$T20-\gamma1.1$ & 4546 & 1.48 & 0.065 $\pm$ 0.006 & 0.81 \\ 
$T20-\gamma1.8$ & 4493 & 1.62 & 0.066 $\pm$ 0.007 & 0.83 \\ 
$T15-\gamma1.3$ & 4245 & 1.55 & 0.067 $\pm$ 0.008 & 0.82 \\   \hline
\end{tabular}
\begin{tablenotes}
			\item[a]  The quoted errors are purely statistical.
\end{tablenotes}
\end{threeparttable}
\label{tab:Gamma-constraint-1}
\end{table}

\begin{table}
\caption{Joint (Flux PDF + Flux PS) $1\sigma$ constraint on $\Gamma_{\rm 12}$ for different thermal history (see Table \ref{tab:thermal-history-parameters-low-z}) for redshift bin II (refer Table \ref{tab:obs-property})}
\begin{threeparttable}
\centering
\begin{tabular}{ccccc}
\hline
 & \multicolumn{2}{c}{$z=0.2$}  & \multicolumn{2}{c}{Joint constraint} \\ \hline
Model & $T_0$ & $\gamma$ & $\Gamma_{\rm 12}$\tnote{a} & $\chi^2_{\rm dof}$ \\ \hline
$T10-\gamma1.1$ & 4326 & 1.53 & 0.105 $\pm$ 0.015 & 1.22 \\ 
$T10-\gamma1.8$ & 4313 & 1.61 & 0.105 $\pm$ 0.015 & 1.22 \\ 
$T20-\gamma1.1$ & 4971 & 1.46 & 0.100 $\pm$ 0.014 & 1.13 \\ 
$T20-\gamma1.8$ & 4889 & 1.62 & 0.100 $\pm$ 0.013 & 1.11 \\ 
$T15-\gamma1.3$ & 4583 & 1.54 & 0.100 $\pm$ 0.015 & 1.18 \\ \hline
\end{tabular}
\begin{tablenotes}
			\item[a]  The quoted errors are purely statistical.
\end{tablenotes}
\end{threeparttable}
\label{tab:Gamma-constraint-2}
\end{table}

\begin{table}
\caption{Joint (Flux PDF + Flux PS) $1\sigma$ constraint on $\Gamma_{\rm 12}$ for different thermal history (see Table \ref{tab:thermal-history-parameters-low-z}) for redshift bin III (refer Table \ref{tab:obs-property})}
\begin{threeparttable}
\centering
\begin{tabular}{ccccc}
\hline
 & \multicolumn{2}{c}{$z=0.3$}  & \multicolumn{2}{c}{Joint constraint} \\ \hline
Model & $T_0$ & $\gamma$ & $\Gamma_{\rm 12}$\tnote{a} & $\chi^2_{\rm dof}$ \\ \hline
$T10-\gamma1.1$ & 4589 & 1.51 & 0.150 $\pm$ 0.021 & 1.05 \\ 
$T10-\gamma1.8$ & 4568 & 1.60 & 0.150 $\pm$ 0.020 & 1.04 \\ 
$T20-\gamma1.1$ & 5383 & 1.44 & 0.145 $\pm$ 0.023 & 0.99 \\ 
$T20-\gamma1.8$ & 5279 & 1.61 & 0.145 $\pm$ 0.022 & 0.99 \\ 
$T15-\gamma1.3$ & 4903 & 1.53 & 0.145 $\pm$ 0.023 & 1.02 \\    \hline
\end{tabular}
\begin{tablenotes}
			\item[a]  The quoted errors are purely statistical.
\end{tablenotes}
\end{threeparttable}
\label{tab:Gamma-constraint-3}
\end{table}

\begin{table}
\caption{Joint (Flux PDF + Flux PS) $1\sigma$ constraint on $\Gamma_{\rm 12}$ for different thermal history (see Table \ref{tab:thermal-history-parameters-low-z}) for redshift bin IV (refer Table \ref{tab:obs-property}). The $\Gamma_{\rm 12}$ constraints are obtained from Ly-$\beta$ contamination analysis (see \S\ref{subsec:Gamma12-constraint})}
\begin{threeparttable}
\centering
\begin{tabular}{ccccc}
\hline
 & \multicolumn{2}{c}{$z=0.4$}  & \multicolumn{2}{c}{Joint constraint} \\ \hline
Model & $T_0$ & $\gamma$ & $\Gamma_{\rm 12}$\tnote{a} & $\chi^2_{\rm dof}$ \\ \hline
$T10-\gamma1.1$ & 4844 & 1.5 & 0.215 $\pm$ 0.025 & 1.09 \\ 
$T10-\gamma1.8$ & 4810 & 1.6 & 0.215 $\pm$ 0.025 & 1.08 \\ 
$T20-\gamma1.1$ & 5811 & 1.42 & 0.205 $\pm$ 0.027 & 1.01 \\ 
$T20-\gamma1.8$ & 5677 & 1.61 & 0.210 $\pm$ 0.030 & 0.96 \\ 
$T15-\gamma1.3$ & 5220 & 1.51 & 0.210 $\pm$ 0.030 & 1.06 \\  \hline
\end{tabular}
\begin{tablenotes}
			\item[a]  The quoted errors are purely statistical.
\end{tablenotes}
\end{threeparttable}
\label{tab:Gamma-constraint-4}
\end{table}

The constraints on $\Gamma_{12}$ can also depend on the cosmological parameters, e.g., any change in $\sigma_{\rm 8}$ can affect the overall normalization of flux PS since Ly-$\alpha$ flux field and the matter density field are anti-correlated. In order to calculate uncertainty in $\Gamma_{12}$, we used Eq. \ref{eq:FGPA} to propagate the error due to uncertainty in cosmological parameters. The uncertainty in cosmological parameters in this work is consistent with \citet{planck2015}. We found that the change in $\Gamma_{12}$ is $\leq 4$ per cent due to uncertainty in $\sigma_8$ (in the range $0.820$ to $0.848$). Similarly, uncertainties in $\Omega_{\rm b} h^2$ and $\Omega_m$  would also affect the constraints on $\Gamma_{12}$ (see Eq. \ref{eq:FGPA}). The combined uncertainty in $\Gamma_{\rm 12}$ due to uncertainty in $\Omega_{\rm b} h^2$ (in the range $0.02184$ to $0.02230$) and $\Omega_{m}$ (in the range $0.297$ to $0.323$) \citep{planck2015} is $\sim 4$ per cent.  Thus we found that the uncertainties in parameters $\Omega_{\rm b} \: h^2$,  $\Omega_m$ and $\sigma_{\rm 8}$ leads to $\leq 10$ per cent change in $\Gamma_{\rm 12}$ measurements. Note that we do not account for the correlation between these parameters. We also found that the derived value of $\Gamma_{\rm 12}$ do not change with change in $n_s$ from $0.96$ to $1.0$.\footnote{We performed a {\sc gadget-2} simulation with $n_s=1.0$ and followed the method described in this paper to constraint $\Gamma_{\rm 12}$.}

One further source of error could come from the cosmic variance. We use a box having a rather moderate size $50 h^{-1}$ cMpc, which could in principle be smaller than the largest voids at $z \sim 0$ \citep[see Fig. 2 in][]{mao2016}. It is thus possible that our measurements of $\Gamma_{12}$ may not be globally representative. To account for the effect, we simulate another box of identical size with identical parameters, however choosing a different set of initial conditions on the density and velocity fields. We perform the same statistical analysis on the second box and find that the constraints on $\Gamma_{12}$ differ by $\leq 3$ per cent. This is consistent with the finding of \citet{smith2011} that simulations do converge for box sizes $\ge 50$ cMpc for the number of particles considered in our simulations.

One possible source of uncertainty in $\Gamma_{\rm HI}$ could come from the metallicity of the IGM. At low-$z$, the IGM is enriched with metals  \citep[typical metallicity $\sim 0.1 Z_{\sun}$ see][]{kulkarni2005,shull2012b}. In presence of metals the cooling rates are enhanced which in turn can affect the $T-\Delta$ relation. To study the effect of metals on $T-\Delta$ relation, we included the cooling due to metals in {\sc cite} using cooling tables given by \citet{wiersma2009}\footnote{http://www.strw.leidenuniv.nl/WSS08/}. 
We enhance the cooling rates (due to metals) while calculating ionization fractions of species and hence temperature evolution. We do not account for the change in density due to metal cooling in {\sc cite}. We assume constant metallicity of $0.1 Z_{\sun}$ from redshift $z_1 = 2.1$ to $z=0.0$.  We found that due to metals the mean IGM temperature $T_0$ decreases by $\sim 10$ per cent whereas $\gamma$ remains same. The fraction of baryons in the diffuse phase increase by $\sim 2.5$ per cent. Since $\Gamma_{\rm 12}$ constraints are weakly dependent on $T_0$ and the fractional change in diffuse phase of baryons is small, the IGM metallicity has little effect on $\Gamma_{\rm 12}$.

The final source of errors on the measured $\Gamma_{12}$ is the uncertainty in the continuum fitted to the observed spectra. As discussed earlier, each observed spectrum is fitted with the continuum by \citet{danforth2016}. However, because of the limited SNR the continuum fitting procedure is prone to uncertainties. 
Conventionally, the observed flux is normalized as $F = F_{\rm unnorm} / F_{\rm cont} $, where $F_{\rm unnorm}$ and $F_{\rm cont}$ are the unnormalized and continuum flux, respectively. If $\delta F_{\rm cont}$ is the value of the uncertainty in the continuum, we can calculate the lower and upper bounds on the normalized flux as  $F_{\rm lb} = F_{\rm unnorm} / (F_{\rm cont} + \delta F_{\rm cont})$ and $F_{\rm ub} = F_{\rm unnorm} / (F_{\rm cont} - \delta F_{\rm cont})$, respectively. Fig. \ref{fig:continuum-flux-pdf-ps-variation} shows the effect of continuum fitting uncertainties on the observed flux PDF (left-hand panel) and flux PS (right-hand panel). When $\delta F_{\rm cont}$ is taken to be $1\sigma$ uncertainty in the continuum, the flux PDF is considerably affected by the continuum uncertainty, whereas the effect on the flux PS is milder and the changes are well within the errorbars. This is the main reason to constrain $\Gamma_{\rm 12}$ from flux PS and flux PDF jointly. In this work, we obtained constraints on $\Gamma_{\rm 12}$ using the three estimates of the transmitted flux $F_{lb}$, $F$,  and $F_{ub}$ and found that the $\Gamma_{\rm 12}$ range changes systematically by $\leq 10$ per cent.

We summarize the total error budget (statistical and systematic) in Table \ref{tab:error-budget}.
The errors on $\Gamma_{12}$ are thus calculated as follows: (i) We first estimate the error through the $\chi^2$ minimization for the fiducial values of $T_0$ and $\gamma$. (ii) We then account for the uncertainties in $T_0$ and $\gamma$ by obtaining constraints for models with extreme values of the two parameters. (iii) We add all statistical uncertainties in quadrature  to account for uncertainties in the cosmological parameters, thermal history and cosmic variance. (iv) We finally add total statistical uncertainty with systematic (from the continuum fitting) uncertainty. The $\Gamma_{\rm 12}$ constraints accounting for statistical and systematic uncertainties are given in Table \ref{tab:error-budget}.

\begin{table*}
\caption{Total error budget for our $\Gamma_{\rm 12}$ measurements at different redshifts}
\begin{threeparttable}
\centering
\begin{tabular}{lccccc}
\hline \hline
Redshift bin $\Rightarrow$ & I &  II & III & IV \\ 
Type of simulated spectra $\Rightarrow$ & Ly-$\alpha$ forest & Ly-$\alpha$ forest &Ly-$\alpha$ forest & Ly-$\alpha$ + Ly-$\beta$ forest\tnote{1}  \\ \hline \hline
Best Fit $\Gamma_{\rm 12}$  & 0.066 & 0.100 & 0.145 & 0.210 \\ \hline
Statistical Uncertainty\tnote{a} & $\pm$ 0.007 & $\pm$ 0.013 & $\pm$ 0.022 & $\pm$ 0.030 \\ 
Cosmological parameters ($\sim 10$ per cent) \tnote{b} & $\pm$ 0.007 & $\pm$ 0.010 & $\pm$ 0.015 & $\pm$ 0.021 \\ 
Cosmic Variance ($\sim 3$ per cent) \tnote{c}  & $\pm$ 0.002  & $\pm$ 0.003 & $\pm$ 0.004 & $\pm$ 0.006 \\ 
Total statistical errors\tnote{d}  & $\pm$ 0.010  & $\pm$ 0.016 & $\pm$ 0.027 & $\pm$ 0.037 \\  \hline
Continuum uncertainty (systematic)\tnote{e}  & $\pm$ 0.005  & $\pm$ 0.005 & $\pm$ 0.010 &$\pm$ 0.015 \\  \hline
Total error\tnote{f}  & $\pm$ 0.015 & $\pm$ 0.021 & $\pm$ 0.037 & $\pm$ 0.052\\  \hline \hline
\end{tabular} 
\begin{tablenotes}
\item[a] The best fit value and statistical uncertainty is given
  for the model $T20-\gamma1.8$ (see Table \ref{tab:thermal-history-parameters-low-z}) since $\chi^2$ is minimum as compared to other models in 3 out of 4 redshift bins.
	       \item[b] Cosmological parameters: Uncertainty due to ($\Omega_{b} h^2$, $\Omega_m$, $\sigma_8$) is ($\sim 2$ per cent, $\sim 2$ per cent, $\sim 4$ per cent) respectively. The correlation between different parameters is not accounted for hence the uncertainty is conservative.
	       \item[c] To account for cosmic variance, we simulated two different boxes with identical cosmological parameters but with different initial conditions. 
            \item[d] All statistical errors are added in quadrature.
            \item[e] This uncertainty arises due to continuum fitting uncertainty. This error is systematic in nature.
            \item[f] Total error is obtained by adding total statistical error with total systematic error.
            	\item[1] Simulated Ly-$\alpha$ forest at $z=0.35$ to $0.45$ is contaminated by Ly-$\beta$ forest in the same wavelength range. The Ly-$\beta$ forest is generated from simulation box at $z=0.6$. 
\end{tablenotes}
\end{threeparttable}
\label{tab:error-budget}
\end{table*}

The redshift bin $z=0.4$ is likely to be contaminated by Ly-$\beta$ lines from H~{\sc i} interlopers \citep{danforth2016}\footnote{Note that one can still identify such lines using Ly-$\gamma$, Ly-$\delta$ transition if the Ly-$\beta$ line is sufficiently saturated. However here we are concerned about the lines which have strong Ly-$\alpha$ and Ly-$\beta$ transition but weak Ly-$\gamma$, Ly-$\delta$ transition.}. The contaminated spectrum will have more absorption hence $\Gamma_{\rm 12}$ would be underpredicted as compared to uncontaminated spectrum. 
To account for this we contaminated the Ly-$\alpha$ forest (at $z=0.4$) with Ly-$\beta$ forest\footnote{We have only one free parameter $\Gamma_{\rm 12}$ at $z=0.4$  denoted by $\Gamma_{\rm 12,0.4}$ in this analysis. For  a given $\Gamma_{\rm 12,0.4}$, we obtained $\Gamma_{\rm 12,0.6}$ at $z=0.6$ by using scaling relation $\Gamma_{\rm 12,0.6} = \Gamma_{\rm 12,norm} (1+z)^{4.4}$ where $\Gamma_{\rm 12,norm} = \Gamma_{12,0.4} / 1.4^{4.4}$ as found by \citet{shull2015}.} from $z=0.6$. We have contaminated the region after accounting for observed QSO emission redshift and avoiding proximity region. Fig. \ref{fig:ly-beta-contamination} shows that $\Gamma_{\rm 12}$ measurement obtained from contaminated spectra are higher as compared to those from uncontaminated spectra (Fig. \ref{fig:result-hst-cos-2015-2} bottom row).
This problem does not arise for other redshift bins because the Ly-$\beta$ lines are identified based on the Ly-$\alpha$ detected in the HST-COS spectrum and we removed these lines, higher H~{\sc i} Ly-series and metal lines in contaminating line removal process as illustrated in Fig. \ref{fig:obs-sim-spectra}. 

 \InputFigCombine{hst-cos-2015-4_Ly_Beta_New.pdf}{160}
{Each panel is same as explained in Fig. \ref{fig:result-hst-cos-2015-2} except that the Ly-$\alpha$ forest in the redshift range $z=0.4 \pm 0.05$ is contaminated by Ly-$\beta$ forest from high redshift ($z=0.6$). Comparison of the bottom row in Fig. \ref{fig:result-hst-cos-2015-2} (without Ly-$\beta$ contamination) with this plot shows that the $\Gamma_{\rm 12}$ constraints are underpredicted (at $z=0.4$) when Ly-$\beta$ contamination is not taken into account.}
{\label{fig:ly-beta-contamination}}

\subsection{Evolution of $\Gamma_{\rm 12}$} 
\label{subsec:Gamma12-evolution}

Having obtained the constraints on $\Gamma_{12}$ at different redshift bins, we can now try to understand its redshift evolution. Left-hand panel in Fig. \ref{fig:gamma-evolution} shows the constraints on $\Gamma_{\rm 12}$ using the combined $\chi^2$ analysis of flux PDF and PS at the four redshift bins. Blue open circle at $z=0.4$ in the left-hand panel shows that the $\Gamma_{\rm 12}$ measurements after Ly-$\beta$ contamination is accounted for properly. We find that there is a clear trend of photoionization rate increasing with increasing redshifts. The best fit values follow $\Gamma_{\rm 12} = 0.040 \pm 0.001 (1+z)^{4.99 \pm 0.12}$.
Our $\Gamma_{\rm 12}$ constraint at $z=0.2$ is also in agreement with those obtained from modeling  the observed metal abundances by \citet{shull2014}. However, $\Gamma_{\rm 12}$ constraint at $z=0.1$ are smaller by factor $\sim 2.7$ as compared to \citet{kollmeier2014}.
 
To compare our results with the UVB models, we calculate the UVB as explained in \citet{khaire2013} and KS15. The UVB estimate depends on the UV emissivity of QSOs and galaxies and the H~{\sc i} column density distribution, $f(N_{\rm HI}, z)$, of the IGM. We use the QSO emissivity from KS15 at 912 ${\rm \AA}$ (their Eq. 6) obtained using the recent QSO luminosity functions. At $z<3$, this $\epsilon_{912}$ is higher upto factor of 2.2 than the one used by previous UVB models such as \citet[][hereafter HM12]{faucher2009,hm12} which they obtained from the compiled luminosity functions of \citet{hopkins2007}. It is because the recent QSO luminosity function compiled by KS15 from \citet{palanque2012,croom2009,schulz2009} have more bright QSOs. The ratio of the $\epsilon_{912}$ from KS15 to the one used by HM12 at $z=0.1,0.5,1.0,1.5,2.0,2.5,3.0$ is $1.16,1.61,2.03,2.11,1.86,1.54,1.28$ respectively (see Fig. 2 of KS15).  The QSO emissivity at $\lambda < 912 \: {\rm \AA}$ depends on the assumed spectral energy distribution (SED) of QSOs. The SED can be approximated by a power-law, $L_{\nu} \propto \nu^{\alpha}$, for $\lambda < 912 \: {\rm \AA}$. We refer to $\alpha$ as UV spectral index. Here, we use $\alpha=-1.4$ from \citet{stevans2014} as well as $\alpha=-1.7$ from \citet{lusso2014}. \citet{stevans2014} obtained it  by stacking the FUV spectra of $159$ QSOs at $0.001 < z < 1.476$ observed with HST-COS which probes rest frame wavelengths down to $475 \: {\rm \AA}$. Whereas, \citet{lusso2014} obtain it by stacking spectra of a sample of $53$ QSOs at $z \sim 2.4$ observed using Wide Field Camera $3$ on HST which probes rest wavelength down to $600 \: {\rm \AA}$. To model the UVB, we take the galaxy emissivity using a fiducial self-consistent combination of star formation rate density (SFRD) and dust attenuation from \citet{ks15a} \citep[as summarized in \S3 of][]{kstp2015} with $f_{\rm esc}$ being a free parameter. In general, the UVB at $z<0.5$ has significant contribution coming from high-$z$ sources upto $z \sim 2 $ due to steep rise in QSO emissivity as well as SFRD with $z$.  Therefore, we need to constrain the $f_{\rm esc}$ at $z<2$. For simplicity, we assume a constant $f_{\rm esc}(z)$ over this redshift range. 

The $f(N_{\rm HI}, z)$ for $\log{N_{\rm HI}}>16$ affects the UVB significantly. It is not well constrained at $z<2$ due to the small number statistics and because one needs to use spectra obtained with space based observatories. Here, we use recently updated $f(N_{\rm HI},z)$ from \citet{inoue2014} which is different than the one used by HM12. For comparison, we calculate UVB using  $f(N_{\rm HI}, z)$ of both \citet{inoue2014} as well as HM12.
 
 The $\Gamma_{\rm HI}(z)$ obtained using these UVB models calculated for two UV spectral indexes and two $f(N_{\rm HI}, z)$ are shown in the right-hand panel of Fig. \ref{fig:gamma-evolution}. Here, we assumed $f_{\rm esc}(z)$=0 for $z<3$ and at $z>3$ $f_{\rm esc}(z)$ is taken from \citet{kstp2015}. As shown in Fig. \ref{fig:gamma-evolution}, all four UVB models are consistent with our low-$z$ $\Gamma_{\rm HI}$ measurements.
As can be seen from the figure, QSOs alone are sufficient to explain the $\Gamma_{\rm HI}$ measurements upto $z=2.5$ (i.e with $f_{\rm esc}(z)=0$). At higher-$z$ a rapid evolution in $f_{\rm esc}$ is needed as shown in \citet{kstp2015}.
We calculate the reduced $\chi^2$ for these UVB models obtained with $f_{\rm esc}=0$. The UVB with $f(N_{\rm HI}, z)$ of HM12 gives reduced $\chi_{\rm dof}^2 = 0.14$ and $1.26$ for $\alpha=-1.4$ and $\alpha=-1.7$, respectively. The UVB with $f(N_{\rm HI}, z)$ of  \citet{inoue2014} gives reduced $\chi_{\rm dof}^2 = 3.06$ and  $0.22$ for $\alpha=-1.4$ and $\alpha=-1.7$, respectively. 

For UVB model obtained using $f(N_{\rm HI}$, $z)$ of  \citet{inoue2014} and $\alpha=-1.7$, we find that $f_{\rm esc}(z<2)=0.008$ is a conservative $3\sigma$ upper limit (with $\Delta \chi^2 \sim 9$). It is consistent with the $3\sigma$ upper limits on average $f_{\rm esc} \leq 0.02$ obtained by stacking samples of galaxies  \citep{siana2010,cowie2009,bridge2010,rutkowski2015}.  In these observations, lowest average mass of galaxies is $\sim 10^{9.3}$ M${_{\sun}}$ \citep{rutkowski2015}. It was believed that the UV escape from the lower mass galaxies could be appreciable at all $z$ \citep{fujita2003,razoumov2006,ferrara2013,wise2014} which may solve the problem of higher $f_{\rm esc}$ \citep[$\sim 0.15$ to $0.2$][]{mitra2015,kstp2015} at $z>6$ required for H~{\sc i} reionization. However our derived 3-$\sigma$ upper limit of $f_{\rm esc}$ is in conjunction with observations of \citet{rutkowski2015}, suggests that galaxies with mass lower than $10^{9.3}$ M$_{\sun}$ may also have very low $f_{\rm esc}$ providing negligible contributions to the UVB at low-$z$.
It is possible that some additional heating as suggested recently by \citet{viel2016} may be present and if included that will reduce the derived $\Gamma_{\rm HI}$ further. This will further strengthen our conclusion that QSOs alone are sufficient to provide necessary H~{\sc i} ionizing photons.

\InputFigCombine{Continuum_Flux_PDF_PS_Variation.pdf}{160}
{\emph{Left} and \emph{right-hand panels} show variation in flux PDF and PS (for observed spectra at $z = 0.2$) with uncertainty in continuum respectively.  $F_{\rm unnorm}$ and $F_{\rm cont}$ are unnormalized and continuum flux respectively. $\delta F_{\rm cont}$ indicates $1 \sigma$ uncertainty in continuum flux. Due to continuum uncertainty the flux PDF is affected more as compared to flux PS}
{\label{fig:continuum-flux-pdf-ps-variation}}

\InputFigCombine{Gamma_12_evolution.pdf}{170}
                {\emph{Left-hand} panel shows the $\Gamma_{\rm 12}$  constraint from joint $\chi^2$ analysis of flux PS and flux PDF. The red diamonds show our $\Gamma_{\rm 12}$ constraints using the simulated Ly-$\alpha$ forest. The last ($z=0.4$) bin is likely to be affected by Ly-$\beta$ forest from H~{\sc i} interlopers at high redshift. The blue open circle corresponds to the $\Gamma_{\rm 12}$ constraint using simulated Ly-$\alpha$ forest contaminated by Ly-$\beta$ forest at $z=0.6$. A best fit power-law to our measurements is also shown with green dashed line.
The scaling relation used by \citet{shull2015} (black dashed line), where they increased the $\Gamma_{\rm 12}$ evolution of HM12 by a factor 2, is also consistent with our measurements. However our $\Gamma_{\rm 12}$ at $z=0.1$ is factor $\sim 2.7$ smaller than \citet{kollmeier2014} (yellow star). 
\emph{Right-hand} panel shows the $\Gamma_{\rm 12}$ evolution from $z = 0$ to $2.5$ from observations and different UVB models.  The cyan and orange shaded regions show evolution of $\Gamma_{\rm 12}$ from KS15 UVB for $f_{\rm esc} = 0$   using \citet{inoue2014} and HM12 cloud distribution respectively. The shaded region accounts for uncertainty in UV spectral index $\alpha = -1.4$ to $-1.7$ at $\lambda < 912$ $\AA$. Our results  (shown by red diamonds) are consistent with $f_{\rm esc} = 0$ for HM12 and \citet{inoue2014} cloud distribution allowing for the UV spectral index uncertainties. 
A constant $f_{\rm esc} = 0$ model (for different cloud distribution and FUV spectral index uncertainty) is sufficient to explain the evolution of  $\Gamma_{\rm 12}$ from $z=0$ to $z=2.5$ (high-$z$ points are taken from \citet{bolton2007} and \citet{becker2013}). All of these predictions use the QSO emissivity of KS15, and no galaxy contribution.}
{\label{fig:gamma-evolution}}

\subsection{Column density distribution function (CDDF)} 
\label{subsec:cdd}
The column density distribution (hereafter CDDF) $f(N_{\rm HI},z)$ is defined as the number of absorbers with column density between $N_{\rm HI}$ to $N_{\rm HI} + d N_{\rm HI}$ and in the redshift range $z$ to $z+dz$.\footnote{Some authors used the absorption path length $X$ instead of redshift $z$. The two quantities are related by $dX = dz\:  (1+z)^2 \:  \frac{H_0}{H(z)}$ see \citep{bahcall1969}.}
Previously \citet{kollmeier2014} and \citet{shull2015} used CDDF to constrain $\Gamma_{\rm HI}$ from Ly-$\alpha$ forest because $f(N_{\rm HI},z) \propto \Gamma_{\rm 12}^{-1}$.
To calculate the CDDF, each absorption line is fitted using Voigt profile with column density ($N_{\rm HI}$), doppler $b$ parameter and line center ($\lambda_c$) as free parameters. 
We have developed our own automatic Voigt profile fitting code for Ly-$\alpha$ forest.
Following \citet{danforth2016}, we first identify the lines using significance level (SL) cutoff ${\rm SL} > 2$ where SL is defined as, 
\begin{equation}
\label{eq:significance-level}
\begin{aligned}
{\rm SL} = \frac{W(\lambda)}{\sigma(\lambda)}
\end{aligned}
\end{equation}
where $W(\lambda)$ and $\sigma(\lambda)$ are equivalent width vector and `line less error' vector.
We then fit all the lines in spectrum simultaneously by allowing parameters to vary in the range $10 \leq \log (N_{\rm HI}) \leq 16.5$, $8 \leq b \leq 150 \: {\rm km \: s}^{-1}$ and $\lambda_c \pm 0.1 \: {\rm  \AA}$.
As a test to our code we automatically fitted the voigt profiles to the Ly-$\alpha$ lines in the observed HST-COS spectra and found within errorbars most of our fitted parameters agree with \citet{danforth2016} parameters.\footnote{\citet{danforth2016} used a semi-automatic technique. For highly saturated lines they fit two or more component and check if fit is improved. The number of such components is small and we are using CDDF for consistency check hence we have not incorporated multiple component fit to saturated lines.} We will present the detail descriptions and comparison  of our code with other codes elsewhere.

Fig. \ref{fig:cdd} shows the observed CDDF (red dashed line) and simulated CDDF (blue dotted line) in the redshift range $0.075 \leq z \leq 0.45$. We assume that the errorbars on observed CDDF are poisson distributed. The simulated CDDF is calculated from mock suite (mock suite consists of $N \times N_{spec}$ spectra). To calculate the CDDF from simulation (blue dotted line), we used best fit $\Gamma_{\rm 12}$ obtained here (which is evolving with redshift) in the given redshift bin (see Table \ref{tab:error-budget}). The shaded region shows the simulated CDDF range obtained by shifting the $\Gamma_{12}$ by $\pm 1 \sigma$ range in $\Gamma_{12}$ (see Table. \ref{tab:error-budget}).
From figure it is clear that within errors the observed CDDF and simulated CDDF are consistent with each other.
The reduced $\chi^2$ between observed CDDF and simulated CDDF is $0.85$ which is also an indication of goodness of fit.
For calculating reduced $\chi^2$ we consider only the error in the simulated CDDF and ignore the measurement errors in the observed CDDF.
Note that our $\Gamma_{\rm 12}$ measurement at $z=0.1$ is a factor $\sim 2.7$ smaller than \citet{kollmeier2014}.

\InputFig{CDD.pdf}{80}
         {Comparison of CDDF measured from observations (red dashed curve) and our simulations (blue dotted curve) using best fitted $\Gamma_{12}(z)$ in the redshift range $0.075 \leq z \leq 0.45$. The shaded region corresponds to 1$\sigma$ range in CDDF (obtained by shifting $\Gamma_{12}$ by $\pm 1\sigma$, see text for more details) that is consistent $(\chi^2_{\rm dof} = 0.85)$ with the 1$\sigma$ constraints on $\Gamma_{\rm 12}$ as a function of $z$.
        }
{\label{fig:cdd}}

\section{Summary}   
\label{sec:summary}
In this work we measure the H~{\sc i} photoionization rate, $\Gamma_{\rm HI}$, at $z\le 0.45$ using a sample of QSO spectra obtained with Cosmic Origins Spectrograph onboard the Hubble Space Telescope  (HST-COS) and hydrodynamical simulations using {\sc gadget-2}.
We developed a new module ``{\sc cite}'' to evolve the temperature of the IGM from high redshift $2.1$ to $0$
in the post-processing step of the {\sc gadget-2} simulation taking into account various photo-heating and radiative cooling processes. 

We compare our results with other low-$z$ simulations in the literature using three predictions. These are,
(i) thermal history parameters: our simulation predicts $T_0 \sim 5000$ K and $\gamma \sim 1.6$ in the redshift range $z=0.1$ to $0.45$. These values are shown to be insensitive to our choice of
$T_0$ and $\gamma$ at an initial redshift, $z_1=2.1$;
(ii) distribution of baryons in phase diagram at $z=0$: We find $\sim 34$ per cent of baryons are in diffuse phase, $\sim 29$ per cent in WHIM, $\sim 18$ per cent in hot halo and $\sim 19$ per cent in condensed phase and 
(iii) the correlation between baryon overdensity $\Delta$ vs H~{\sc i} column density, $N_{\rm HI}$, in the redshift range $0.2 < z < 0.3$: we find $\Delta = 34.8 \pm 5.9 \: ( N_{\rm HI} / 10^{14})^{0.770 \pm 0.022}$.
We show that all these predictions compare well with those of low-$z$ simulations in the literature that include different feedback processes at varied levels.
Feedback processes such and galactic winds or AGN feedback are not incorporated in our simulations.
However, as shown by \citet{shull2015} these processes are not expected to severely influence $\Gamma_{\rm HI}$ constraints. Our method is computationally less expensive and allowed us to quantify various statistical and systematic uncertainties associated with the $\Gamma_{\rm HI}$ measurements.

For fair comparison, we mimic the simulated Ly-$\alpha$ forest as close to observations as possible in terms of  SNR, resolution and line spread function. The spectra generated using our method are remarkably similar to the observed spectra. We use two statistics that avoid voigt profile decomposition of the
spectra (i.e., flux PDF and PS)  and $\chi^2$ minimization
using appropriate covariance matrices to compare the observations
with the model predictions. Using simulated data we show that these two statistics are good in recovering
the $\Gamma_{\rm HI}$.
The main results of our work are as follows,
\begin{enumerate}
\item We measured $\Gamma_{\rm HI}$ in four different redshift bins (of $\Delta z~=~0.1$) centered at $z=0.1125,0.2,0.3,0.4$ using joint constraints from the two statistics mentioned above. We estimated the associated errors by varying thermal history parameters, cosmological parameters and continuum fitted to the observed spectrum. Due to limited wavelength range covered in the HST-COS spectrum used in this study, the $\Gamma_{\rm HI}$ measurement for the highest redshift bin (i.e $z=0.4$) is likely to be affected by the contamination of Ly-$\beta$ forest absorption from higher-$z$. We contaminated our simulated Ly-$\alpha$ forest at $z=0.4$ by Ly-$\beta$ forest from $z=0.6$ and corrected for the effect of Ly-$\beta$ contamination in our  $\Gamma_{\rm HI}$ measurement for this $z$ bin. The measured $\Gamma_{\rm 12}$ values at  redshift bins $z=0.1125,0.2,0.3,0.4$ are $0.066 \pm 0.015,~0.100 \pm 0.021,~0.145 \pm 0.037,~0.210 \pm 0.052$, respectively.
\item Our final quoted errors in the $\Gamma_{\rm HI}$ measurements include possible uncertainties coming from the statistical uncertainty\footnote{The percentage values given in parenthesis are quoted for redshift bin IV i.e., $z=0.4$ (see Table \ref{tab:error-budget})} ($\sim 14$ per cent), cosmic variance ($\sim 3$ per cent), cosmological parameters uncertainty ($\sim 10$ per cent) and continuum uncertainty (systematic uncertainty $\sim 7$ per cent). Uncertainty in $\Gamma_{\rm HI}$ due to uncertainty in thermal history parameters, over the range considered here, is small and within statistical uncertainty.
\item As expected based on UVB models, even in the small redshift range covered in our study the measured $\Gamma_{\rm HI}$ shows a rapid evolution with $z$. We fit the redshift evolution of $\Gamma_{\rm 12}$ as $\Gamma_{\rm 12} = 0.040 \pm 0.001 \: (1+z)^{4.99 \pm 0.12}$ at $0.075 \le z \le 0.45$.

\item The $\Gamma_{\rm HI}(z)$ obtained here are consistent with the measurement of \citet{shull2015} however our $\Gamma_{\rm HI}$ measurement at $z = 0.1$ is factor $\sim 2.7$ smaller than \citet{kollmeier2014}. Note these two earlier measurements used H~{\sc i} column density distribution to constrain $\Gamma_{\rm HI}(z)$. As a consistency check, we show the H~{\sc i} column density distribution predicted by our simulations, for the best fit value of  $\Gamma_{\rm 12}(z)$ we have obtained using the statistics (flux PDF and PS), matches well with the observed distribution. 
  
\item The $\Gamma_{\rm HI}$ measurement at any $z_1$ depends on the emissivities of the ionizing sources at $z\ge z_1$ and Lyman continuum opacity of the IGM.  We considered the updated emissivities of QSOs and galaxies
  (with $f_{\rm esc}$ as a free parameter) and two different H~{\sc i} column density distribution as a function of $z$ obtained by HM12 and \citet{inoue2014} and obtained  $\Gamma_{\rm HI}$ using KS15 UVB code. We find that for, both H~{\sc i} distributions, our derived  $\Gamma_{\rm HI}(z)$ is consistent with being contributed only by QSOs. This is true even if we allow for variations in the UV spectral index of QSOs. We also find the maximum $3\sigma$ upper limit on $f_{\rm esc}$ at $z< 2$, allowing for uncertainty in FUV spectral index and cloud distribution $f(N_{\rm HI},z$) of \citet{inoue2014}, is $0.008$. This is consistent with $3\sigma$ upper limits on average $f_{\rm esc}$  (i.e $\le 0.02$) obtained by stacking samples of galaxies probing average galaxy mass $M\ge 10^{9.3} M_\odot$.

  Our measurements suggest that the contribution of low mass galaxies to average $f_{\rm esc}$  will also be small.{\it Our study confirms that there is no crisis at low redshift in accounting for the observed Lyman continuum photons using standard known luminous astronomical sources}. Thus our $\Gamma_{\rm HI}(z)$ measurement can in turn be used to place a strong constraint on the contributions of  decaying dark matter to the low-$z$ UVB.
\end{enumerate}
\section*{Acknowledgment}
We would like to thank Aseem Paranjape, Sowgat Muzahid and Charles Danforth for useful discussion on statistics and observations. All the computations were performed using the PERSEUS cluster at IUCAA and the HPC cluster at NCRA.
\bibliographystyle{mnras}
\bibliography{UV_background}

\section*{Appendix}
We tabulate the flux PDF, the PS and mean flux decrement ($\rm DA$) obtained from the observational data used in this paper, and also the values corresponding to our best-fit model. The errors on flux PDF and flux PS are obtained from the simulated mock samples.
\paragraph*{Mean flux decrement:} Following the standard definition the mean flux decrement along a sightline is given by, 
\begin{equation}
{\rm DA} = \langle 1 - {\rm e}^{-\tau} \rangle
\end{equation}
where ${\rm e}^{-\tau}$ is normalized flux and angle brackets represent average along the wavelength.
The total mean flux decrement for the $N$ sightlines is given by,
\begin{equation}\label{eq:mean-DA}
{\rm DA_{sample}} = \frac{1}{N} \sum \limits_{i=1}^{N} {\rm DA}_i
\end{equation}
where ${\rm DA}_i$ is mean flux decrement along $i^{\rm th}$ sightline.
The variance (${\rm \sigma^2_{sample}}$) of the total mean flux decrement for the $N$ sightlines is given by,
\begin{equation}\label{eq:sigma-DA}
{\rm \sigma^2_{sample}} = \frac{1}{N} \sum \limits_{i=1}^{N} ({\rm DA}_i - {\rm DA_{sample}})^2
\end{equation}
Table \ref{tab:mean-flux-decrement} shows the observed (${\rm DA_{data}}$) and best fit flux decrement (${\rm DA_{model}}$) from our simulation for different redshift bins (calculated using Eq. \ref{eq:mean-DA}).
The uncertainty in ${\rm DA_{data}}$ is calculated using Eq. \ref{eq:sigma-DA} whereas the uncertainty in ${\rm DA_{model}}$ corresponds to $1\sigma$ uncertainty in best fit $\Gamma_{\rm 12}$. The uncertainty in ${\rm DA_{data}}$ due to continuum fitting uncertainty is not accounted for. Note that for $\chi^2$ analysis, we used flux PDF in the range $F \leq 0.85$.

\begin{table*}
\caption{The observed and best fit $\Gamma_{\rm 12}$ (see Table \ref{tab:error-budget}) flux PDF from our simulation for redshift bins. The errorbars on model flux PDF indicate diagonal terms of the covariance matrix. The covariance matrix can be available on request.}
\begin{tabular}{ccccccccc}
\hline \hline
Spectra $\rightarrow$ & \multicolumn{2}{c}{Ly-$\alpha$ forest} & \multicolumn{2}{c}{Ly-$\alpha$ forest} & \multicolumn{2}{c}{Ly-$\alpha$ forest} & \multicolumn{2}{c}{Ly-$\alpha$ + Ly-$\beta$ forest} \\ \hline
 & $z=0.1125$ &  $z=0.1125$ &  $z=0.2$ &  $z=0.2$ &  $z=0.3$ &  $z=0.3$ &  $z=0.4$ & $z=0.4$ \\ \hline
F & ${\rm P}_{\rm data}$ & ${\rm P}_{\rm model} \pm {\rm dP}_{\rm model}$ & ${\rm P}_{\rm data}$ & ${\rm P}_{\rm model} \pm {\rm dP}_{\rm model}$ & ${\rm P}_{\rm data}$ & ${\rm P}_{\rm model} \pm {\rm dP}_{\rm model}$ & ${\rm P}_{\rm data}$ & ${\rm P}_{\rm model} \pm {\rm dP}_{\rm model}$ \\ \hline \hline
0 & $0.027$ & 0.046 $\pm$ 0.011 & $0.022$ & 0.049 $\pm$ 0.016 & $0.019$ & 0.054 $\pm$ 0.016 & $0.031$ & 0.074 $\pm$ 0.022 \\ 
0.1 & $0.028$ & 0.028 $\pm$ 0.004 & $0.025$ & 0.030 $\pm$ 0.006 & $0.027$ & 0.032 $\pm$ 0.006 & $0.034$ & 0.042 $\pm$ 0.008 \\ 
0.2 & $0.018$ & 0.020 $\pm$ 0.003 & $0.022$ & 0.021 $\pm$ 0.005 & $0.020$ & 0.023 $\pm$ 0.004 & $0.030$ & 0.031 $\pm$ 0.006 \\ 
0.3 & $0.018$ & 0.020 $\pm$ 0.003 & $0.023$ & 0.022 $\pm$ 0.004 & $0.022$ & 0.024 $\pm$ 0.004 & $0.032$ & 0.033 $\pm$ 0.006 \\ 
0.4 & $0.021$ & 0.024 $\pm$ 0.003 & $0.023$ & 0.026 $\pm$ 0.005 & $0.030$ & 0.030 $\pm$ 0.005 & $0.035$ & 0.039 $\pm$ 0.006 \\ 
0.5 & $0.030$ & 0.031 $\pm$ 0.003 & $0.029$ & 0.033 $\pm$ 0.005 & $0.047$ & 0.046 $\pm$ 0.008 & $0.052$ & 0.056 $\pm$ 0.007 \\ 
0.6 & $0.047$ & 0.048 $\pm$ 0.004 & $0.053$ & 0.048 $\pm$ 0.006 & $0.077$ & 0.087 $\pm$ 0.017 & $0.100$ & 0.103 $\pm$ 0.013 \\ 
0.7 & $0.097$ & 0.106 $\pm$ 0.010 & $0.092$ & 0.094 $\pm$ 0.011 & $0.182$ & 0.212 $\pm$ 0.038 & $0.229$ & 0.271 $\pm$ 0.039 \\ 
0.8 & $0.386$ & 0.429 $\pm$ 0.037 & $0.319$ & 0.362 $\pm$ 0.050 & $0.626$ & 0.674 $\pm$ 0.073 & $0.745$ & 0.828 $\pm$ 0.086 \\ 
0.9 & $2.079$ & 2.229 $\pm$ 0.057 & $2.093$ & 2.250 $\pm$ 0.083 & $2.282$ & 2.379 $\pm$ 0.052 & $2.334$ & 2.458 $\pm$ 0.051 \\ 
1 & $7.247$ & 7.018 $\pm$ 0.094 & $7.300$ & 7.066 $\pm$ 0.134 & $6.668$ & 6.439 $\pm$ 0.112 & $6.375$ & 6.065 $\pm$ 0.116 \\ \hline \hline
\end{tabular}
\label{tab:flux-pdf-data}
\end{table*}

\begin{table*}
\caption{The observed and best fit $\Gamma_{\rm 12}$ (see Table \ref{tab:error-budget}) flux PS from our simulation for redshift bins. The errorbars on model flux PS indicate diagonal terms of the covariance matrix. The covariance matrix can be available on request. The wavenumber $k$ is expressed in units of $h \; {\rm cMpc}^{-1}$.}
\begin{tabular}{ccccccccc}
\hline \hline
Spectra $\rightarrow$ & \multicolumn{2}{c}{Ly-$\alpha$ forest} & \multicolumn{2}{c}{Ly-$\alpha$ forest} & \multicolumn{2}{c}{Ly-$\alpha$ forest} & \multicolumn{2}{c}{Ly-$\alpha$ + Ly-$\beta$ forest} \\ \hline
 & $z=0.1125$ &  $z=0.1125$ &  $z=0.2$ &  $z=0.2$ &  $z=0.3$ &  $z=0.3$ &  $z=0.4$ & $z=0.4$ \\ \hline
 & Data & Model & Data & Model & Data & Model & Data & Model \\ 
$\log k$ & $\log P_F$ & $\log P_F \pm$ d$\log P_F$ & $\log P_F$ & $\log P_F \pm$ d$\log P_F$ & $\log P_F$ & $\log P_F \pm$ d$\log P_F$ & $\log P_F$ & $\log P_F \pm$ d$\log P_F$ \\ 
\hline \hline
$-0.601$ & $-4.332$ & -4.338 $\pm$ 0.145 & $-4.340$ & -4.308 $\pm$ 0.149 & $-3.971$ & -3.995 $\pm$ 0.204 & $-3.706$ & -3.652 $\pm$ 0.162 \\ 
$-0.425$ & $-4.297$ & -4.360 $\pm$ 0.142 & $-4.312$ & -4.339 $\pm$ 0.153 & $-4.107$ & -4.025 $\pm$ 0.198 & $-3.717$ & -3.685 $\pm$ 0.158 \\ 
$-0.300$ & $-4.337$ & -4.394 $\pm$ 0.143 & $-4.361$ & -4.364 $\pm$ 0.151 & $-4.082$ & -4.045 $\pm$ 0.189 & $-3.741$ & -3.716 $\pm$ 0.158 \\ 
$-0.203$ & $-4.391$ & -4.418 $\pm$ 0.140 & $-4.369$ & -4.395 $\pm$ 0.144 & $-4.102$ & -4.082 $\pm$ 0.187 & $-3.813$ & -3.749 $\pm$ 0.155 \\ 
$-0.124$ & $-4.408$ & -4.459 $\pm$ 0.133 & $-4.433$ & -4.436 $\pm$ 0.145 & $-4.150$ & -4.119 $\pm$ 0.187 & $-3.718$ & -3.799 $\pm$ 0.151 \\ 
$-0.057$ & $-4.421$ & -4.490 $\pm$ 0.132 & $-4.413$ & -4.466 $\pm$ 0.141 & $-4.120$ & -4.151 $\pm$ 0.180 & $-3.713$ & -3.835 $\pm$ 0.148 \\ 
$0.047$ & $-4.477$ & -4.543 $\pm$ 0.123 & $-4.446$ & -4.521 $\pm$ 0.131 & $-4.234$ & -4.212 $\pm$ 0.169 & $-3.857$ & -3.894 $\pm$ 0.135 \\ 
$0.142$ & $-4.547$ & -4.632 $\pm$ 0.119 & $-4.518$ & -4.616 $\pm$ 0.123 & $-4.214$ & -4.302 $\pm$ 0.157 & $-3.915$ & -4.002 $\pm$ 0.128 \\ 
$0.236$ & $-4.589$ & -4.732 $\pm$ 0.114 & $-4.633$ & -4.724 $\pm$ 0.120 & $-4.323$ & -4.411 $\pm$ 0.146 & $-4.012$ & -4.127 $\pm$ 0.117 \\ 
$0.331$ & $-4.675$ & -4.843 $\pm$ 0.108 & $-4.745$ & -4.842 $\pm$ 0.111 & $-4.469$ & -4.521 $\pm$ 0.130 & $-4.207$ & -4.257 $\pm$ 0.110 \\ 
$0.425$ & $-4.839$ & -4.957 $\pm$ 0.100 & $-4.897$ & -4.963 $\pm$ 0.101 & $-4.560$ & -4.628 $\pm$ 0.123 & $-4.309$ & -4.386 $\pm$ 0.099 \\ 
$0.520$ & $-4.977$ & -5.072 $\pm$ 0.092 & $-5.064$ & -5.087 $\pm$ 0.095 & $-4.756$ & -4.741 $\pm$ 0.109 & $-4.509$ & -4.517 $\pm$ 0.089 \\ 
$0.615$ & $-5.154$ & -5.202 $\pm$ 0.083 & $-5.254$ & -5.224 $\pm$ 0.085 & $-4.864$ & -4.863 $\pm$ 0.098 & $-4.658$ & -4.657 $\pm$ 0.081 \\ 
$0.709$ & $-5.299$ & -5.333 $\pm$ 0.075 & $-5.401$ & -5.364 $\pm$ 0.077 & $-5.013$ & -4.976 $\pm$ 0.089 & $-4.794$ & -4.791 $\pm$ 0.071 \\ 
$0.804$ & $-5.431$ & -5.456 $\pm$ 0.065 & $-5.546$ & -5.496 $\pm$ 0.067 & $-5.161$ & -5.073 $\pm$ 0.077 & $-4.936$ & -4.906 $\pm$ 0.063 \\ \hline \hline
\end{tabular}
\label{tab:flux-ps-data}
\end{table*}

\begin{table*}
\caption{The observed (${\rm DA_{data}}$) and best fit flux decrement (${\rm DA_{model}}$) from our simulation for different redshift bins. }
\begin{threeparttable}
\centering
\begin{tabular}{lccccc}
\hline \hline
Redshift bin $\Rightarrow$ & I &  II & III & IV \\ 
Type of simulated spectra $\Rightarrow$ & Ly-$\alpha$ forest & Ly-$\alpha$ forest &Ly-$\alpha$ forest & Ly-$\alpha$ + Ly-$\beta$ forest\tnote{a}  \\ \hline \hline
Best Fit $\Gamma_{\rm 12}$  & 0.066 $\pm$ 0.015 & 0.100 $\pm$ 0.021 & 0.145 $\pm$ 0.037 & 0.210 $\pm$ 0.052 \\ \hline
${\rm DA_{data}}$\tnote{b} & 0.021 $\pm$ 0.002 & 0.024 $\pm$ 0.002 & 0.025 $\pm$ 0.002 & 0.031 $\pm$ 0.003 \\ 
${\rm DA_{model}}$\tnote{c} & 0.028 $\pm$ 0.005 & 0.030 $\pm$ 0.004 & 0.032 $\pm$ 0.006 & 0.033 $\pm$ 0.006 \\ \hline \hline
\end{tabular} 
\begin{tablenotes}
            	\item[a] Simulated Ly-$\alpha$ forest at $z=0.35$ to $0.45$ is contaminated by Ly-$\beta$ forest in the same wavelength range. The Ly-$\beta$ forest is generated from simulation box at $z=0.6$.
            \item[b] The uncertainty in ${\rm DA_{data}}$ is calculated using Eq. \ref{eq:sigma-DA}. However, the uncertainty in ${\rm DA_{data}}$ due to continuum fitting uncertainty is not accounted for. Note that for $\chi^2$ analysis, we used flux PDF in the range $F \leq 0.85$.
            	\item[c] The uncertainty in ${\rm DA_{model}}$ corresponds to $1\sigma$ uncertainty in best fit $\Gamma_{\rm 12}$.
\end{tablenotes}
\end{threeparttable}
\label{tab:mean-flux-decrement}
\end{table*}







\bsp	
\label{lastpage}
\end{document}